\documentclass[english,twocolumn,pra,showpacs,aps]{revtex4}

\usepackage[T1]{fontenc}
\usepackage[utf8]{inputenc}
\usepackage{graphicx}
\usepackage{amssymb,bm}
\usepackage{amsmath}
\usepackage{babel}

\usepackage{color}
\usepackage{comment}

\usepackage[driverfallback=dvipdfm]{hyperref}
\hypersetup{
	colorlinks,
	citecolor=blue,
	filecolor=blue,
	linkcolor=blue,
	urlcolor=blue
}

\newcommand{\bea}{\begin{eqnarray}}

\newcommand{\eea}{\end{eqnarray}}

\newcommand{\beq}{\begin{equation}}

\newcommand{\eeq}{\end{equation}}

\def \bR{{\bf R}}



\begin{document}

\title{Bose-Einstein condensation in a frustrated triangular optical lattice}

\author{Peter Janzen$^{1}$, Wen-Min Huang$^{2,}$\footnote{Email:wenmin@phys.nchu.edu.tw}, L. Mathey$^{1,3,}$\footnote{Email:lmathey@physnet.uni-hamburg.de}}
\affiliation{
$^{1}$Zentrum f\"ur Optische Quantentechnologien and Institut f\"ur Laserphysik, Universit\"at Hamburg, 22761 Hamburg, Germany\\
$^{2}$Department of Physics, National Chung-Hsing University, Taichung 40227, Taiwan\\
$^{3}$The Hamburg Centre for Ultrafast Imaging, Luruper Chaussee 149, Hamburg 22761, Germany
}

\begin{abstract}
The recent experimental condensation of ultracold atoms in a triangular optical lattice with a negative effective tunneling parameter paves the way to study frustrated systems in a controlled environment.
Here, we explore the critical behavior of the chiral phase transition in such a frustrated lattice in three dimensions.
  We represent the low-energy action of the lattice system as 
  a two-component Bose gas corresponding to the two  minima of the dispersion. 
   The contact repulsion between the bosons separates into intra- and inter-component interactions, referred to as $V_{0}$ and $V_{12}$, respectively. 
  We first employ a Huang-Yang-Luttinger approximation of the free energy.  For $V_{12}/V_{0} = 2$, which corresponds to the bare interaction, this approach suggests a first order phase transition, at which both the U$(1)$ symmetry of condensation and the $\mathbb{Z}_2$ symmetry of the emergent chiral order are broken simultaneously.   
%  As a result, the weak contact repulsion between bosons is separated into intra- and inter-component interactions, referred to as $V_{0}$ and $V_{12}$, respectively. In a Huang-Yang-Luttinger approach, we first show that the critical behavior of a thermally driven phase transition is determined by the ratio $V_{12}/V_0$. When $V_{12}/V_0<1$, two U$(1)$ symmetries are broken, while the $\mathbb{Z}_2$ symmetry is preserved. If $V_{12}/V_0>1$, bosons tend to condense at one energetic minimum, thus breaking one U$(1)$ and the $\mathbb{Z}_2$ symmetry. In addition, the free energy shows a continuous phase transition of the $\mathbb{Z}_2$ symmetry breaking in the regime $1<V_{12}/V_{0}\lesssim1.5$. However, when $V_{12}/V_{0}\gtrsim1.5$, a discontinuous phase transition is observed. 
  Furthermore, we perform a  renormalization group calculation at one-loop order. 
  We demonstrate that the coupling regime $0<V_{12}/V_0\leq1$ shares the critical behavior of the Heisenberg fixed point at $V_{12}/V_{0}=1$. For  $V_{12}/V_0>1$ we show that $V_{0}$ flows to a negative value, while $V_{12}$ increases and remains positive.
  This results in a breakdown of the effective quartic field theory due to a cubic anisotropy, and again  suggests  a discontinuous phase transition. 
   \end{abstract}

\date{\today}

\pacs{67.85.Hj, 03.75.Mn, 64.60.ae, 75.10.Hk}
% Bose-Einstein condensates, 67.85.Hj
% multicomponent and spinor condensates, 03.75.Mn
% renormalization-group theory in, 64.60.ae
% classical spin models, 75.10.Hk

\maketitle

%%%%%%%%%%%%%%%%%%%%%%%%%
% Introduction  %%%%%%%%%%%%%%%%%%
\section{Introduction}

The seminal studies \cite{Anderson95,Davis95,Bradley95} reported the observation  
 of Bose-Einstein condensation (BEC) in time-of-flight images, which revealed that ultracold bosons accumulate near the minimum of the dispersion relation and condense below a critical temperature. 
This second order phase transition is the quintessential example for U$(1)$ symmetry breaking, described by complex $\phi^{4}$ theory. 
 Subsequently, numerous studies have addressed the question if and how BEC can be achieved,  by including symmetries apart from the conventional  U$(1)$ symmetry \cite{Lewenstein07,Bloch08}. 
Populating bosons in spatially anisotropic orbitals of an optical lattice, for instance, has been proposed to create exotic superfluid orders \cite{Browaeys05,Isacsson05,Liu06,Kuklov06,muller07,Lim08,Sarma08,Wu09,Sarma11,Li11,Cai11,Hemmerich11,Panahi12,Liu12}. 
 In the experiment reported in \cite{Wirth11,Lewenstein11}, a long-lived metastable state of ultracold bosons in the $p$-band was indeed realized and displays condensation involving two different momenta, visible in time-of-flight measurements. Both experimental work \cite{Hemmerich13} as well as theoretical analysis \cite{Liu06,Wu09} suggest that the two momentum states form a superposition with an imaginary relative phase, in which the atoms condense, leading to  superfluid order with broken time-reversal symmetry. By exploiting a matter wave heterodyning technique, the breaking of time-reversal symmetry via the spontaneous formation of chiral order in this $p$-band bipartite optical lattice is revealed without ambiguity \cite{ Kock14}.

Another  method to create unconventional BEC is to load bosons into the Floquet states of a shaken optical lattice  \cite{Eckardt05,Gemelke,Lignier07,Zenesini09,Chin13,Chin15}. 
In such a shaken triangular lattice the effective tunneling parameters can be tuned to  negative values  \cite{Lewenstein10,Struck11}. 
  The bosonic atoms condense at the minima of the effective dispersion at non-zero momenta. 
   This was used  to perform simulations of frustrated classical magnetism in \cite{Struck11}. 
  Furthermore, complex-valued effective tunneling parameters can be created, which correspond to 
artificial gauge fields \cite{Garcia12,Struck12}.
 These artificial gauge fields couple to an emergent, chiral order parameter, as demonstrated in Ref.\cite{Struck13}. 
 
\begin{figure}
\includegraphics[height=8.4cm]{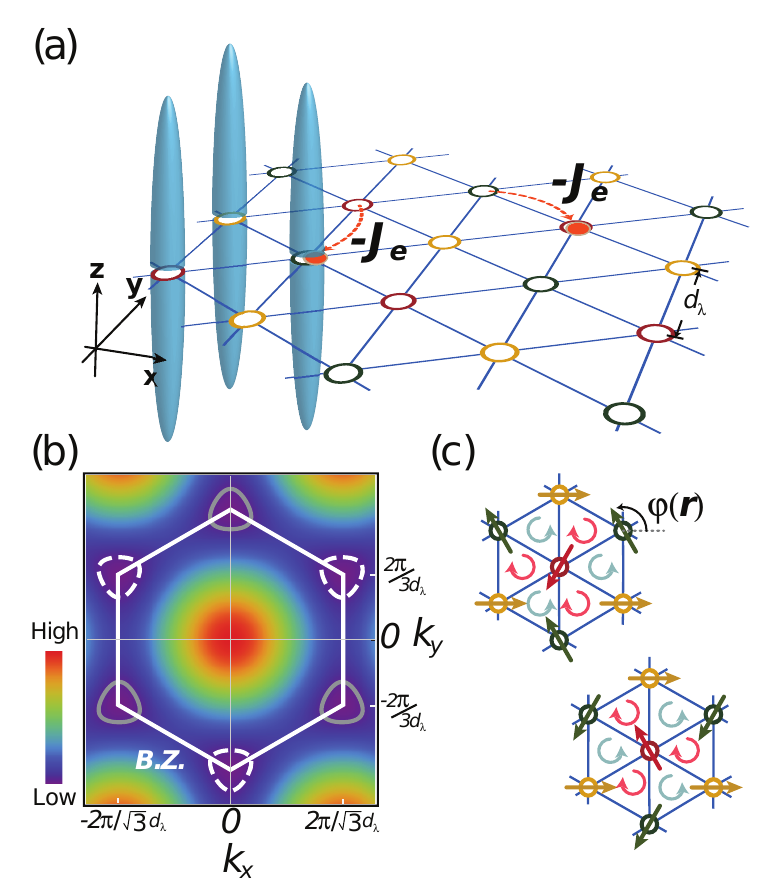}
\caption{(a) Sketch  of the two-dimensional triangular lattice of weakly coupled 1d systems.  $J_e$ is the effective tunneling parameter, with $J_{e}<0$. The bosons move freely in the $z$-direction, and in an optical lattice in the $x-y$ plane. (b) Contour plot of the energy dispersion in the first Brillouin zone (BZ) for the triangular lattice model, in the $k_{x} - k_{y}$ plane, based on Eq. \ref{eqn:dispersion}. Contours near the two distinct minima are depicted as  dashed and solid lines. (c) Phase configurations of the two condensates at momenta $\textbf{k}_{\rm{1}}$ and $\textbf{k}_{\rm{2}}$ are equivalent to the   spin fields $\bm{s}(\bm{r})=(\cos\varphi(\bm{r}),\sin\varphi(\bm{r}))$. }
\label{fig:spins}
\end{figure}

The frustrated bosonic system studied in Ref.\cite{Struck13}, as sketched in Fig.~\ref{fig:spins} (a), is related to a classical XY model in the triangular lattice by ignoring phase fluctuations along the z-axis. The underlying physics is to represent the bosonic fields $\langle\psi(\bm{r})\rangle=\sqrt{n(\bm{r})}e^{i\varphi(\bm{r})}$ in the phase-density representation, thus the phase $\varphi(\bm{r})$ is mapped onto a classical XY spin, $\bm{s}(\bm{r})=(\cos\varphi(\bm{r}),\sin\varphi(\bm{r}))$ \cite{Wen}. For a negative tunneling parameter $J_e<0$ in this approximation results in  antiferromagnetic spin coupling between neighboring sites \cite{Struck13,Becker10}. This two-dimensional (2D) antiferromagnetic XY model is frustrated due to its triangular geometry, and has two degenerate ground states, discriminated by their chiral order, as illustrated in Fig.~\ref{fig:spins} (c),  \cite{Choi,Yosefin,Hasenbusch05}. Thus, in addition to breaking the U$(1)$ symmetry of the system,  spontaneous breaking of an emergent, chiral $\mathbb{Z}_2$ symmetry occurs \cite{Yosefin,Hasenbusch05,Altman14}.

%

%%%%%%%%

%Though the critical behavior of the fully frustrated XY model in 2D has been studied extensively in the past \cite{Yosefin,Hasenbusch05,Altman14}, we  explore the critical  behavior of an interacting Bose gas  in the geometry shown in Fig. \ref{fig:spins} (a) in this paper

 %Though the fully frustrated XY model in 2D has been studied extensively in the past \cite{Yosefin,Hasenbusch05,Altman14}, the critical behavior of the $\mathbb{Z}_2$ symmetry breaking, condensation at only one of the two degenerate energetic minima, accompanied with the U$(1)$ symmetry breaking in this bosonic system are still lost in literature.
 
In this paper, we explore the critical behavior of an interacting Bose gas, realized in Ref.\cite{Becker10,Struck11,Struck13}. In the $xy$-plane, as shown in Fig. \ref{fig:spins} (a), the bosons move in a triangular optical lattice with a real-valued, the negative effective tunneling parameter $J_e<0$, while moving freely along the $z$-direction. We therefore include the three dimensional character of the system, and the density fluctuations. We develop a field theoretical description of the system, which has the form of a two-component Bose gas. The two species describe the bosons near the two minima of the dispersion, as shown in Fig.~\ref{fig:spins} (b). In this description, the U$(1)$ symmetry is broken if one of the two species is condensed to the nearby minima. Moreover, the $Z_2$ symmetry breaking is analogous to condensation at only one of the two degenerate energetic minima~\cite{Struck13,Becker10}. The weak contact repulsion of the bosons is also decomposed into the interaction within one component, $V_{0}$, and the interaction between these two components, $V_{12}$, in the two-component Bose gas.

By regarding these interactions as two independent parameters in a Huang-Yang-Luttinger (HYL) approach \cite{Huang57}, we find that the critical behavior of the $U(1) \times U(1) \times \mathbb{Z}_2$ symmetry breaking is controlled by the ratio of these two interactions in the mean-field scheme. The HYL approximation predicts that  the $\mathbb{Z}_2$ symmetry breaking is a continuous, second-order phase transition for $1< V_{12}/V_0\lesssim1.5$. However, the $\mathbb{Z}_2$ symmetry breaking becomes discontinuous, first-order, for $V_{12}/V_0\gtrsim1.5$. For $V_{12}/V_0<1$, the two $U$(1) symmetries are spontaneously broken as the temperature is decreased, while the $\mathbb{Z}_2$ symmetry is preserved.

To further study the phase diagram beyond the mean-field scheme, we perform a one-loop renormalization group (RG) calculation, resulting in a flow of the chemical potential and the coupling constants. 
 This calculation clarifies that there are two regimes: 
For  $0<V_{12}/V_0\leq1$, $V_{12}$ and $V_{0}$ flow towards a Heisenberg fixed point, at which $V_{12}=V_{0}$, describing a second order phase transition that differs from the condensation transition of a single-component system.  
For $V_{12}/V_0>1$, however,  $V_0$ flows towards a negative  value, while $V_{12}$ increases and remains positive. This result implies the breakdown of the quartic theory, and higher-order terms, for instance a three-body interaction, need to be included. It also indicates that a discontinuous phase transition occurs \cite{Domany}, due to a cubic anisotropy. 
 Thus, we conclude that the critical behavior in the regime $V_{12}/V_0>1$ is of first order, improving on the HYL approximation. We conclude that for the bare interaction of the system, the ratio $V_{12}/V_0 = 2$, suggests that the transition is first order.

This paper is organized as follows: in Sec.~\ref{tri} we develop the field theoretic description of the lattice model. In Sec.~\ref{yhl} we study the free energy based on the Huang-Yang-Luttinger approximaiton and investigate the phase diagram for different ratios of $V_{12}/V_0$. In Sec.~\ref{rg}, we study the critical behavior in the framework of a one-loop RG calculation,  and in Sec.~\ref{conc} we conclude.

%%%%%%%%%%%%%%%%%%%%%%%%
%   Two-component Bose-Hubbard model in a triangular optical lattice    %%%%%%%%%%%%%%
\section{Effective field theory}\label{tri}
The system, as depicted in Fig. \ref{fig:spins} (a), is described by  the Hamiltonian $H=H_{0}+H_I$, with 
\begin{eqnarray}\label{h0}
\nonumber &&\hspace{-0.5cm}H_{0} = \int dz~
\Bigg\{ \sum_{\bm{r}} \psi^{\dagger}(\bm{r},z)\left(\frac{-\hbar^{2}\partial^2_z}{2m_0}-\mu_{3D}\right) \psi(\bm{r},z)\\
&&\hspace{1.5cm}+|J_e| \sum_{\langle\bm{r},\bm{r}'\rangle} \left[\psi^{\dagger}(\bm{r},z)\psi(\bm{r}',z)+\rm{h.c.}\right]\Bigg\},
\end{eqnarray}
and the interaction term
\begin{eqnarray}\label{hi}
H_{I} = \frac{U}{2}\sum_{\bm{r}}\int dz~
\psi^{\dagger}(\bm{r},z)\psi^{\dagger}(\bm{r},z)\psi(\bm{r},z)\psi(\bm{r},z).
\end{eqnarray}
$\langle\bm{r},\bm{r}'\rangle$ represents nearest-neighbor pairs of
sites and $m_0$ is the  mass of the atoms. $\mu_{3D}$ is the chemical potential, and $U$ denotes the magnitude of the repulsive contact interaction. 
 The frustration of the system is due to the negative value of the tunneling parameter $J_{e}$. 
  Neglecting the interaction term and the chemical potential, we obtain the dispersion relation  
\begin{eqnarray}
\nonumber &\hspace{-3mm}\varepsilon (\textbf{k})=\left|J_e\right|\Big[2 \cos( d_{\lambda} k_{y})+2 \cos\left( d_{\lambda}  \sqrt{3} k_{x}/2 - d_{\lambda}k_{y}/2\right) \\
&+2 \cos\left( d_{\lambda}  \sqrt{3} k_{x}/2+d_{\lambda}k_{y}/2\right)\Big]+ \hbar^{2} k_{z}^{2} / (2 m_{0}) ,
\label{eqn:dispersion}
\end{eqnarray}
with the lattice spacing $d_{\lambda}=2\lambda_L/3$, and $\lambda_L$ being the wavelength of the laser creating the 2D lattice potential \cite{Struck13}. The  dispersion is shown in Fig.~\ref{fig:spins} (b). It displays two energetically degenerate minima, located at the two distinct momenta $\textbf{k}_{\rm{1/2}}=\left( 2 \pi/(\sqrt{3} d_{\lambda}) , \pm 2 \pi/(3 d_{\lambda}) \right)$, at the boundary of the first Brillouin zone. The classical spins $\bm{s}(\bm{r})$ depicted in Fig.~\ref{fig:spins} (c) correspond to the real and imaginary part of the plane waves $\exp( i \textbf{k}_{\rm{1}} \bm{r})$ and $\exp( i \textbf{k}_{\rm{2}} \bm{r})$.

The hidden $U(1) \times U(1) \times \mathbb{Z}_2$ symmetry of the system emerges as follows. At low temperatures, the bosons will accumulate near the  minima. If the bosons condense at one of the two minima, one $U(1)$ symmetry is broken due to condensation, and the $\mathbb{Z}_2$ symmetry,  which corresponds to the density imbalance between the minima. However, if the bosons condense at both minima with equal density, the $\mathbb{Z}_2$ symmetry is preserved, while both $U(1)$ symmetries are broken.

We now derive the low-energy field theoretical description.  At low temperatures, bosons will be distributed around the minima $\textbf{k}_{\rm{1}}$ and $\textbf{k}_{\rm{2}}$. Therefore, we decompose the bosonic field in momentum space as
\begin{eqnarray}
\nonumber\psi(\bm{r},z)&=&\frac{1}{\sqrt{N}}\sum_{\bm{k}}e^{i\bm{k}\cdot\bm{r}}\psi(\bm{k},z)\\
\nonumber&\simeq&\frac{\mathcal{A}_{\lambda}^2}{\sqrt{N}}\hspace{-0.1cm}\sum_{j=1,2}e^{i\bm{k}_j\cdot\bm{r}}\hspace{-0.1cm}\int_{|\bm{q}_j|<\Lambda_q}\hspace{-0.2cm}\frac{d^2\bm{q}}{4\pi^2}~e^{i\bm{q}_j\cdot\bm{r}}\phi_j(\bm{q}_j,z)\\
&=&\mathcal{A}_{\lambda}\sum_{j=1,2}\phi_j(\bm{r},z)~e^{i\bm{k}_j\cdot\bm{r}},
\end{eqnarray}
where $N$ is the total number of sites in the xy-plane, and $\bm{q}_{j}=\bm{k}-\bm{k}_j$. We define the slowly varying  fields via $\phi_j(\bm{r},z)\equiv\frac{\mathcal{A}_{\lambda}}{\sqrt{N}}\int_{|\bm{q}_j|<\Lambda_q}\frac{d^2\bm{q}}{4\pi^2}~e^{i\bm{q}_j\cdot\bm{r}}\phi_j(\bm{q}_j,z)$ with $\phi_j(\bm{q}_j,z)\equiv\psi_j(\bm{k}_j+\bm{q}_j,z)/\mathcal{A}_{\lambda}$, where $\Lambda_q$ is the momentum cutoff, and $\mathcal{A}_{\lambda}=\sqrt{3}d_{\lambda}^2/2$ is the area in the xy-plane that corresponds to a single site, when mapped on a continuum description. By expressing the Hamiltonian of Eq.~(\ref{h0}) in terms of the fields $\phi_j(\bm{r},z)$, we obtain 
\begin{eqnarray}\label{h0e}
H_0^{\rm eff}=\hspace{-0.15cm}\int\hspace{-0.1cm} d^3\bm{R}\hspace{-0.1cm}\sum_{j=1,2}\sum_{a}\phi_j^{\dag}(\bm{r},z)\left\{-\frac{\hbar^{2}\partial^2_a}{2m_a}-\mu\right\}\phi_{j}(\bm{r},z).
\end{eqnarray}
Here, $a=x,y,z$, $m_x=m_y=m_{J}=\hbar^{2} /(3 d_{\lambda}^{2} |J_e|)$, $\bm{R}=(\bm{r},z)=(x,y,z)$, $m_{z} = m_{0}\simeq0.021m_J$,  and $\mu$ is the chemical potential of the Bose gas in the continuum description.  The fields $\phi_j(\bm{r},z)$ have  a quadratic dispersion relation centered at momentum $(\bm{k}_j,k_z=0)$, respectively, 
\begin{eqnarray}
\varepsilon_{\rm eff} (\bm{q}_{j},k_z) = \frac{\hbar^{2} q_{jx}^{2}}{2 m_{J}} + \frac{\hbar^{2} q_{jy}^{2}}{2 m_{J}} + \frac{\hbar^{2} k_{z}^{2}}{2 m_{z}} ,
\end{eqnarray}
with $\bm{q}_{j}=(q_{jx},q_{jy})=\bm{k}-\bm{k}_j$ and $j=1,2$. Expressed in terms of $\phi_j(\bm{r},z)$, the interaction in Eq.~(\ref{hi}) is 
\begin{eqnarray}\label{hie}
\nonumber &&\hspace{-0.8cm}H_I^{\rm eff}=\int d^3\bm{R}~\Bigg\{\frac{V_{0}}{2}\sum_{j=1,2}\phi_j^{\dag}(\bm{R})\phi_j^{\dag}(\bm{R})\phi_j(\bm{R})\phi_j(\bm{R})\\
&&\hspace{2.cm}+V_{12}~\phi_1^{\dag}(\bm{R})\phi_{2}^{\dag}(\bm{R})\phi_{2}(\bm{R})\phi_1(\bm{R})\Big\},
\end{eqnarray}
where $V_{0}$ is the intra-component interaction and $V_{12}$ is the inter-component one. It is noticed that the effective interaction is forbidden to exchange species because of the momentum conservation of the original Hamiltonian, Eq.~(\ref{hi}). Secondly, the bare magnitudes of these parameters are $V_{0}=U/A_{\lambda}$ and $V_{12}=2U/A_{\lambda}$, so in particular we have $V_{12} = 2 V_{0}$.
However, under the RG flow that we derive below, these parameters will flow independently. This forces us to regard $V_{0}$ and $V_{12}$ as two independent variables throughout, because their ratio is not protected by a symmetry. Furthermore, our analysis covers the full universality class of this type, not necessarily limited to the original system on a triangular lattice. The total effective Hamiltonian $H^{\rm eff}=H_0^{\rm eff}+H_I^{\rm eff}$ is a two-component complex $\phi^{4}$-theory, with a single chemical potential $\mu$, because only the total density of the bosons is conserved.

%%%%%%%%%%%%%%%%%%%%%%%%%%%%%%%
%   Huang-Yang-Luttinger approach    %%%%%%%%%%%%%%
\section{Huang-Yang-Luttinger approach}\label{yhl}
 As a first insight into the critical behavior of the system, we compute the free energy of the effective Hamiltonian within the Huang-Yang-Luttinger approximation \cite{Huang57}.
 %we study phase transitions in different ratios of $V_{12}$/$V_{0}$. 
In this approach, the free energy $A$ is expanded in the interaction strength. Thus, the zero order term $A_{0}$ is the free energy of 
 two ideal Bose gases. 
%Therefore, we can first write down the free energy of two free Bose gases in thermal equilibrium. 

To formulate this quantity, we define the average thermal de Broglie wavelength 
\begin{eqnarray}
%\lambda_{0} & =&2 \pi \hbar / \sqrt{2 \pi m_{0} k_{B} T} , \\
%\lambda_{J} & =& 2 \pi \hbar / \sqrt{2 \pi m_{J} k_{B} T} , \\
\lambdabar & =& (\lambda_{0} \lambda_{J}^{2})^{1/3},
\end{eqnarray}
with $\lambda_{0} \equiv 2 \pi \hbar / \sqrt{2 \pi m_{0} k_{B} T} $ and $\lambda_{J} \equiv 2 \pi \hbar / \sqrt{2 \pi m_{J} k_{B} T}$. 
 The density of excited atoms for each of the minima is 
\begin{eqnarray}\label{z}
n_{e,j} = \frac{1}{\lambdabar^{3}} \mathcal{G}_{3/2} (z_{j}) ,
\end{eqnarray}
with $j=1,2$, and $\mathcal{G}_{p} = \sum_{l=1}^{\infty} z^{l} / l^{p}$ being the standard Bose function. $z_{j} = \exp (\mu_{j}/k_{B} T)$ is the fugacity and is computed by the inverse function of Eq.~(\ref{z}) in the free energy below. Here we formally introduce two chemical potentials $\mu_{j}$, but for the result further down we preserve only the total density. 
  The total density of this system is $n_{3D}=n_1+n_2$, where $n_j$ is the density for the $j$th component of the system, and is fixed in free energy computation below. The density of the ground states is defined as $n_{0,j} = n_{j} - n_{e,j}$, which leads to a condensate density of  $n_{0,j}=n_j-\frac{1}{\lambdabar^{3}} \mathcal{G}_{3/2} (1)$, if $n_j> \frac{1}{\lambdabar^{3}}\mathcal{G}_{3/2} (z_{j}=1)$. Therefore, the free energy is given by $A_{0} = A_{0,1} + A_{0,2}$ with
\begin{eqnarray}
\frac{A_{0,i}}{\mathcal{V}} =
\begin{cases}
-\frac{k_{B} T}{\lambdabar^{3}} \mathcal{G}_{5/2} (z_{i}) + n_{i} k_{B} T \ln (z_{i}) & \text{if}\ z_{i}<1 , \\
-\frac{k_{B} T}{\lambdabar^{3}} \mathcal{G}_{5/2} (1) & \text{if}\ z_{i}=1 ,
\end{cases}
\end{eqnarray}
and $\mathcal{V}$ being the volume of the system. 

We now compute the next order of  the free energy $A_{I}$ which is due to and linear in the interaction term (\ref{hie}). We approximate $\langle \phi^{\dag}_j(\bm{q}_j=0,k_z=0) \phi_j(\bm{q}_j=0,k_z=0)\rangle=n_{0,j}$ and $\langle \phi^{\dag}_j(\bm{q}_j,k_z) \phi_j(\bm{q}_j,k_z)\rangle=n_{j}$, according to the HYL approximation. Therefore, the first order contribution to the free energy  is 
\begin{eqnarray}\label{Ain}
\frac{A_{I}}{\mathcal{V}} = \frac{V_{0}}{2} \left[ 2 n_{1}^{2} + 2 n_{2}^{2} - n_{0,1}^{2} - n_{0,2}^{2} \right] + V_{12} n_{1} n_{2}.
\end{eqnarray}
We explore the properties of this approximation of the free energy $A_{0}+A_{I}$ for the 
 experimental parameters of Ref.~\cite{Struck13}, for various ratios of $V_{12}/V_0$ and temperatures, and with fixed total density $n_{3D}$. 
    In units of the lattice constant $d_{\lambda}=553$nm and the effective tunneling parameter  $|J_e|=k_B\times0.26$nK, the interaction strength is given by $V_0\simeq37.78 |J_e|d_{\lambda}^3$, and we have a density of  $n_{\rm 3D}\simeq2.88d_{\lambda}^{-3}$ of $^{87}$Rb atoms.

\begin{figure}
\includegraphics[height=8.3cm,angle=90]{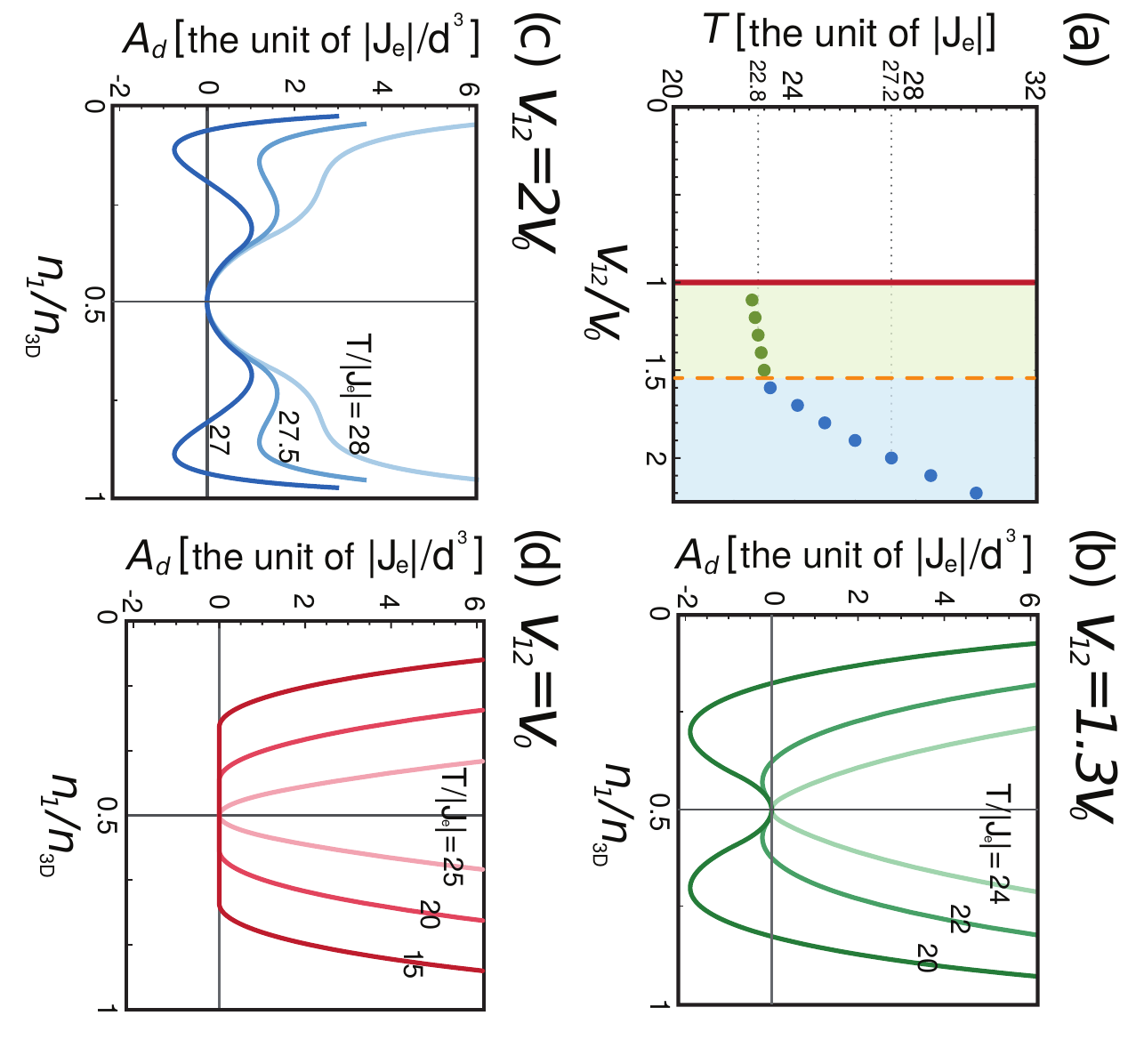}
\caption{(a) The critical temperature of the system, as a function of  $\gamma_{12}=V_{12}/V_0$. 
   For $V_{12}<V_{0}$, the transition corresponds to breaking both $U(1)$ symmetries, while the $\mathbb{Z}_{2}$ symmetry is preserved. 
 For $V_{12}>V_{0}$, one of the $U(1)$ symmetries and the $\mathbb{Z}_{2}$ symmetry are broken. 
    The green shaded regime corresponds to a second order transition of $\mathbb{Z}_{2}$ symmetry breaking, the blue-shaded regime to a first order transition. The critical temperatures for $V_{12}=1.3V_0$ and $V_{12}=2V_0$ are $22.8$ and $27.2$ in the unit of $|J_e|$ respectively.
     (b) In the regime of $1< \gamma_{12}\lesssim1.5$, the two minima of $A_d(n_1)$ in Eq.~(\ref{Ad}) evolve away from $n_1=n_{\rm 3D}/2$ gradually, indicating a continuous phase transition (green lines). The critical temperature for $V_{12}=1.3V_0$ is $22.8$ marked in the panel (a) For $1.5\lesssim \gamma_{12}<2.5$, the minima of $A_d(n_1)$ suddenly occur away from the symmetry point $n_1=n_{\rm 3D}/2$, corresponding to a discontinuous phase transition (blue lines). (d) At $\gamma_{12}=1$, a plateau grows at $A_d(n_1)=0$ by decreasing temperature.}
\label{fig:MFA}
\end{figure}

 At low temperatures, the density of the two condensates $n_{0,j}$ becomes non-zero, which implies  a U$(1)$ symmetry breaking. 
 To analyze the critical behavior of the chiral phase transition, we consider the magnitude of the free energy relative to the $ \mathbb{Z}_2$-symmetric state with $n_{1} = n_{2} = n_{3D}/2$
\begin{eqnarray}\label{Ad}
&&\hspace{-1.cm}\nonumber A_d(n_1)=\frac{1}{\mathcal{V}}\Big[A_0(n_1)+A_I(n_1)\\
&&\hspace{1.5cm}-A_0(n_{3D}/2)-A_I(n_{3D}/2)\Big].
\end{eqnarray}
 If $A_d(n_1)$ becomes negative as the temperature is lowered, it indicates that the $\mathbb{Z}_2$ symmetry is broken. Furthermore, where the new minima emerge indicates if the phase transition is first or second order.

We find that for $V_{12}/V_0<1$ the minimum of $A_d(n_1)$ is always located at $n_1=n_{\rm 3D}/2$, indicating that the $\mathbb{Z}_{2}$ symmetry is preserved. However, for $V_{12}/V_0>1$, we find that  the $\mathbb{Z}_{2}$ symmetry breaks, as new minima appear away from the symmetric point. The temperature at which these minima occur is the critical temperature of the $\mathbb{Z}_{2}$ symmetry breaking. 
It  is plotted against the ratio $\gamma_{12}=V_{12}/V_0$  in Fig.~\ref{fig:MFA} (a).
We also find that the order of the  $\mathbb{Z}_{2}$ symmetry breaking changes as $V_{12}/V_0$ is varied. 
 As shown in Fig.~\ref{fig:MFA} (b), the two minima of $A_d(n_1)$ emerge continuously at $n_1=n_{\rm 3D}/2$, indicating a second order phase transition in the regime of $1<V_{12}/V_{0}\lesssim1.5$. In this regime, as shown in Fig.~\ref{fig:MFA} (a), the critical temperature of the $\mathbb{Z}_{2}$ symmetry breaking only increases weakly with increasing $V_{12}/V_{0}$. 
  However, for $1.5\lesssim \gamma_{12}<2.5$, the minima of $A_d(n_1)$ emerge away  from the symmetric point at $n_1=n_{\rm 3D}/2$, as demonstrated in Fig.~\ref{fig:MFA} (c), indicating a discontinuous phase transition. The critical temperature rapidly increases with increasing $V_{12}/V_{0}$, as illustrated in Fig.~\ref{fig:MFA} (a). 
 % We notice that we constrain our analysis within the regime of $V_{12}/V_{0}<2.5$, because above this regime the total interaction is far beyond the weak coupling limit. 
For $V_{12}/V_0=1$, $A_d$ develops a plateau at $A_d=0$, below the critical temperature, as shown in Fig.~\ref{fig:MFA} (d). This reflects the emergent $SU(2)$ symmetry at this interaction strength, which corresponds to a Heisenberg fixed point, as we discuss below.

We again emphasize that this analysis suggests a first transition of the system, because the bare value of the interaction is $V_{12}/V_0=2$. Furthermore, as we discuss in the next section, our RG analysis suggests that the first order regime extends throughout the entire regime $V_{12}>V_{0}$.

\begin{figure}
\begin{center}
\includegraphics[height=3.85cm]{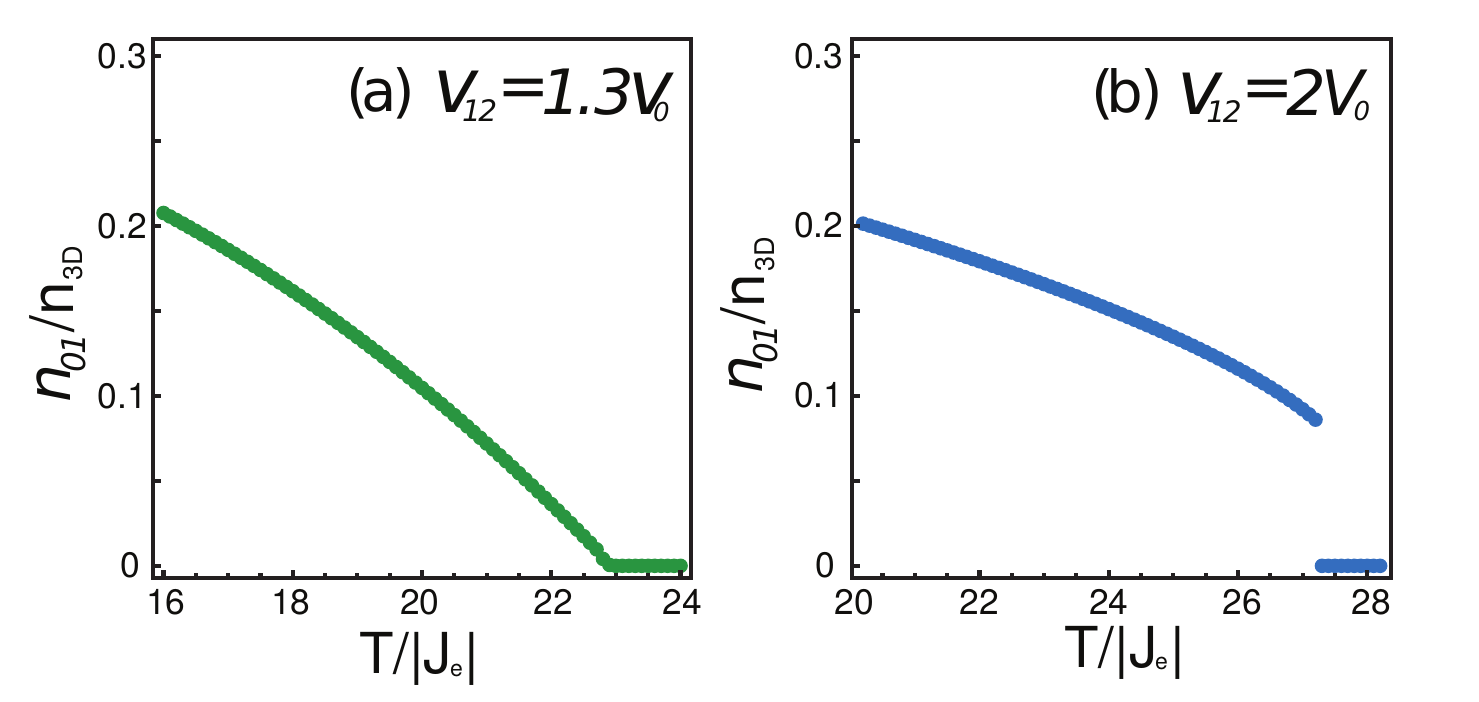}
\caption{The condensate density $n_{0,1}$ as a function of temperature for (a) $V_{12}/V_{0}=1.3$ and (b) $V_{12}/V_{0}=2$  in the HYL approximation. }
\label{fig2-2}
\end{center}
\end{figure}  

We note that the $\mathbb{Z}_{2}$ and the $U(1)$ symmetry breaking occur at the same temperature, within the HYL approximation. This occurs because the condensate density is responsible for generating the two minima in the free energy. We demonstrate this computing the condensate fraction $n_{0,1}$ versus temperature in Fig.~\ref{fig2-2} (a) for $V_{12}/V_{0}=1.3$ and (b) $V_{12}/V_{0}=2$ respectively. The critical temperature of the $U(1)$ symmetry breaking is the same as the temperature at which $n_{0,1}$ becomes non-zero. The first- and second-order phase transitions of the simultaneous $\mathbb{Z}_{2}$ and $U(1)$ symmetry breaking are indicated by the discontinuous and continuous change of the condensate density in Fig~\ref{fig2-2}.

%In summary, we draw different critical behaviors of the $\mathbb{Z}_{2}$ symmetry breaking versus $V_{12}/V_{0}$ as different background colors in Fig.~\ref{fig:MFA}a. In the next section, we will study the phase diagram with the renormalization group method. 

%%%%%%%%%%%%%%%%%%%%%%%%%%%%%%%%%
%   Renormalization Group approach    %%%%%%%%%%%%%%
\section{Renormalization Group approach}\label{rg}
\subsection{Renormalization Group flow}
To study the phase diagram systematically, we perform a renormalization group calculation  at one-loop order in the weak-coupling regime\cite{Stoof09}. 
We write the partition function as a  path integral: 
\begin{eqnarray}
\mathcal{Z} = \int \mathcal{D}\left[ \phi^{*}, \phi \right] e^{-S[\phi^{*}, \phi] / \hbar} ,
\end{eqnarray}
with the action
\begin{eqnarray}
S[\phi^{*}, \phi] \equiv \int_{0}^{\hbar\beta} d \tau \sum_{j=1,2} \Big[  \phi_{j}^{*}\left(\hbar \partial_{\tau}\right) \phi_{j} + H(\phi_{j}^{*}, \phi_{j}) \Big],
\end{eqnarray}
and $H=H_0^{\rm eff}+H_I^{\rm eff}$ in the coherent-state representation of the Hamiltonian $H$. We consider the classical limit of the effective action $S[\phi^{*}, \phi]$, i.e. we ignore the dependence of the fields on the Matsubara time $\tau$,  $\phi_{j}(\tau) = \phi_{j}$. This is equivalent to only taking the $\omega=0$ Matsubara frequency into account \cite{Stoof09}. In this approximation, the partition function simplifies to
\begin{eqnarray}
\mathcal{Z} = \int \mathcal{D}\left[ \phi^{*}, \phi \right] e^{- \beta F_{L}[\phi^{*}, \phi]} ,
\end{eqnarray}
where $F_{L}[\phi^{*}, \phi]=F_{L,0}+F_{L,I}$ is the Landau free energy functional with 
\begin{eqnarray}
F_{L,0} = 
 \int d^3 \bm{R} \sum_{j=1,2}\Bigg[ \frac{\hbar^{2}}{2m^*}\left| \nabla \phi_{j} \right|^{2} -\mu \left| \phi_{j} \right|^{2} \Bigg],
 \label{eqn:free-energy-0}
\end{eqnarray}
and the interaction
\begin{eqnarray}
F_{L,I} = 
 \hspace{-0.1cm}\int \hspace{-0.1cm}d^3 \bm{R} \Bigg[\frac{V_0}{2} \left( \left| \phi_{1} \right|^{4} + \left| \phi_{2} \right|^{4} \right) + V_{12} \left| \phi_{1} \right|^{2} \left| \phi_{2} \right|^{2}\Bigg].
 \label{eqn:free-energy-I}
\end{eqnarray}
We rescale the momenta in the $x$- and $y$-direction by $\sqrt{3 d^{2} |J_e| m_{z} / \hbar^{2}}$ to arrive at a spatially homogeneous expression, with the effective mass $m^*=(m_J^2m_z)^{1/3}$ and $m_{xy}=m_z=m^*\simeq0.276m_J$. To perform the RG transformation, we introduce an energy cutoff of this system $\varepsilon_{\Lambda} = \hbar^{2} \Lambda^{2} / (2m^*)$ with the momentum cutoff $\Lambda$, which sets the maximum energy scale of the field theory description. A physical choice for the energy cut-off is the bandwidth of the dispersion in the x-y plane, which is determined by $J_{e}$.

We note that in the following we assume that the low-energy regime of the driven system is approximately given by $\exp(- \beta H)$. The assumption is based on the high-frequency lattice shaking in our system, where the lattice shaking frequency is $\omega=2\pi\times 2.8 {\rm kHz}$ and the bare hopping amplitude is $J_{bare}=4\times 10^{-3} {\rm E_{rev}}$ i.e. $\hbar\omega/J_{bare}\sim200$~\cite{Struck13}. Though from the eigenstate thermalization hypothesis all the Floquet eigenstates are indistinguishable from the infinite-temperature state or a completely random state at long-time limit~\cite{Breuer,Russomanno,Hone,Iadecola,Rigol,Moessner,Shirai,Liu,Abanin}, in this study we are interested in the quasi-stationary(or metastable) state during high-frequency driving. Recent theoretical study shows for an isolated Floquet systems in high-frequency driving with local interactions, a nonintegrated system as we considered here, heating will be exponentially slow. It implies that prethermalized Floquet many-body phases, though metastable, will be very long-lived~\cite{Abanin2}. On the other hand, because the energy is almost conserved up to timescale $e^{\mathcal{O}(\hbar\omega/J_{bare})}$, the system will first reaches a quasi-stationary state with a {\em finite temperature}~\cite{Saito,Kuwahara}. Thus, thermal equilibrium due to the approximative character of the Floquet Hamiltonian is well-approximated for the large frequencies and the experimentally relevant times considered. Such that, the ensemble average of the periodically driven isolated systems in the intermediate timescale is approximated by the ensemble average of the quasi-stationary state(or Floquet Hamiltonian) in this study. The above concept is also applied in recent theoretical studies~\cite{Demler,Bukov}.

In an RG step, we split the fields $\phi_{j}$ into a low-energy and a high-energy part. The high-energy degrees of freedom are integrated out to obtain a renormalized free energy functional with renormalized chemical potential $\mu$ and coupling constants $V_0$ and $V_{12}$, 
%The corrections to the coupling constants are arranged as a perturbation series in $V_0$ and $V_{12}$. 
 which we compute at one-loop order. Iterating this RG step results in a set of flow equations, given below. 
 The derivation of the flow equations using the $\varepsilon$-expansion is discussed in Appendix \ref{RGderivation}.

We define the dimensionless variables $\mu_{\Lambda} \equiv \frac{1}{\varepsilon_{\Lambda}} \mu$, 
 $g_{0} \equiv \frac{\Lambda^{3}}{2 \pi^{2} \varepsilon_{\Lambda}} V_{0}$, 
 $g_{12} \equiv \frac{\Lambda^{3}}{2 \pi^{2} \varepsilon_{\Lambda}} V_{12}$, 
 $T_{\Lambda} \equiv \frac{1}{\varepsilon_{\Lambda}} k_{B} T$.
  We note that by construction all of these dimensionless quantities are small compared to $1$, because they are divided by the high energy cut-off $\varepsilon_{\Lambda}$.
    For this reason, we terminate the flow if one of these parameters reaches unity, as described below.   
In terms of these variables, the flow equations to linear order in $\varepsilon = 4 - d$ are
\begin{align}
\dfrac{d \mu_{\Lambda}}{dl} &= 2 \mu_{\Lambda} - 2 g_{0} T_{\Lambda} (1 + \mu_{\Lambda}) - g_{12} T_{\Lambda} (1 + \mu_{\Lambda}) , \label{eqn:flow-final1-alt} \\
\dfrac{d g_{0}}{dl} & = \varepsilon g_{0} - 5 g_{0}^{2} T_{\Lambda} - g_{12}^{2} T_{\Lambda} , \label{eqn:flow-final2-alt} \\
\dfrac{d g_{12}}{dl} & = \varepsilon g_{12} - 2 g_{12}^{2} T_{\Lambda} - 4 g_{0} g_{12} T_{\Lambda} . \label{eqn:flow-final3-alt}
\end{align}
where $l = \ln(\Lambda/\Lambda_b)$ is the logarithm of the ratio between
the initial momentum cutoff $\Lambda$ and the running cutoff  $\Lambda_b$. 
%All the dimensionless variables are less than one even during RG transformations, since $\varepsilon_{\Lambda}$ is the maximum energy scale of our theory, as mentioned above. 
We note that by setting $g_{12}=0$, these flow equations reduce to the two standard RG equations of the one-component, complex $\phi^4$ theory. 

\begin{figure}
\includegraphics[height=6.5cm]{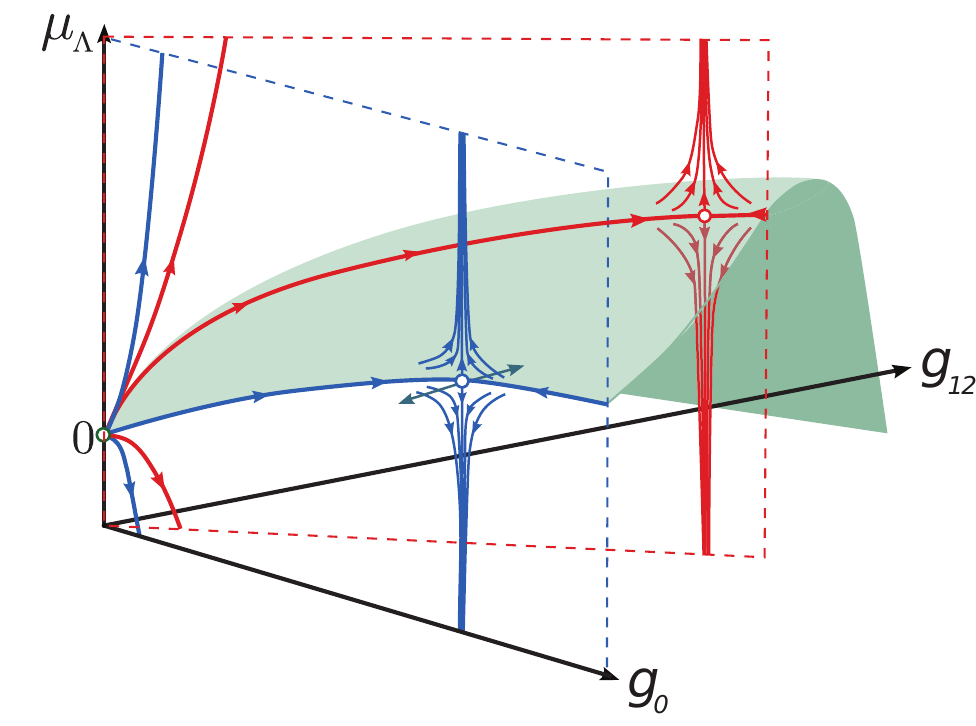}
\caption{Qualitative sketch of the flow diagram in the space spanned by $g_0$, $g_{12}$ and $\mu_\Lambda$. We illustrate the $g_{12}$=$0$ plane (blue) and the $g_{0}$=$g_{12}$ plane (red) with flow trajectories in the vicinity of the three fixed points. The boundary of the thermal and condensate phase is sketched as a transparent green surface.}
\label{fig:flow-3d}
\end{figure}

To identify the fixed points of the flow, we set the Eqs.~(\ref{eqn:flow-final1-alt}), (\ref{eqn:flow-final2-alt}) and (\ref{eqn:flow-final3-alt}) to zero. This results in three fixed points, which are given by 
\begin{eqnarray}
&& \mu_{\Lambda}^{\ast} = 0, \quad g_{0}^{\ast} =0, \quad g_{12}^{\ast} = 0 , \label{eqn:fixed-trivial} \\
&& \mu_{\Lambda}^{\ast} = \frac{\varepsilon}{5}, \quad  g_{0}^{\ast} = \frac{\varepsilon}{5 T_{\Lambda}}, \quad  g_{12}^{\ast} = 0 , \label{eqn:fixed-old} \\
&& \mu_{\Lambda}^{\ast} = \frac{\varepsilon}{4}, \quad  g_{0}^{\ast} = \frac{\varepsilon}{6 T_{\Lambda}}, \quad  g_{12}^{\ast} =\frac{\varepsilon}{6 T_{\Lambda}},\label{eqn:fixed-new}
\end{eqnarray}
to linear order in $\varepsilon$.
 For these points, which are illustrated in Fig.~\ref{fig:flow-3d}, the variables remain unchanged under the RG flow. 
   We investigate the flow further by expanding the flow equations to first order around the fixed points (see Appendix~\ref{flow-first-order}).
%%
%By expanding the flow equations to first order around the fixed points (see Appendix~\ref{flow-first-order}), the behavior of the flow in the vicinity of them can be explored. 
This expansion shows that the non-interacting fixed point~(\ref{eqn:fixed-trivial}) is unstable in all directions. 
 In the vicinity of the second fixed point~(\ref{eqn:fixed-old}),  we recover the flow in the $g_{12}=0$ plane \cite{Stoof09}, which includes one relevant and one irrelevant direction. In addition we find one more relevant direction, which drives the system away from the $g_{12}=0$ plane.
  In Fig.~\ref{fig:flow-3d}, we qualitatively sketch the flow diagram in the vicinity of three fixed points. The irrelevant and relevant directions obtained in the fixed-point analysis correspond to the flow toward the fixed points and away from it, respectively. On the $g_{12}=0$ plane, the line through $(\mu_{\Lambda},g_0)=(0,0)$ and $(\varepsilon/5 , \varepsilon / (5 T_{\Lambda}) )$ corresponds to the boundary between the thermal phase and the condensate phase. 
    Below and above this line, the 
    chemical potential  flows to large negative and positive values, corresponding to $U(1)$ symmetry preserving and breaking, respectively. For non-zero values of $g_{12}$, we find that $g_{12}$ increases under the RG transformation. This indicates that the inter-component interaction is always relevant in a two-component Bose gas. 

The flow behavior in the vicinity of the third fixed point~(\ref{eqn:fixed-new}) is also depicted in Fig.~\ref{fig:flow-3d}. 
  We refer to this fixed point as a Heisenberg fixed point, because it displays $SU(2)$ symmetry. 
  The fixed-point analysis shows that there is  one irrelevant direction along  the $g_{0}=g_{12}$ plane, which is similar to the behavior close to the second fixed point on the $g_{12}=0$ plane. In addition, there is a relevant direction, that drives the chemical potential to $\pm \infty$. 
 The line through $(\mu_{\Lambda},g_0,g_{12})=(0,0,0)$ and $(\varepsilon/4,\varepsilon/(6T_{\Lambda}),\varepsilon/(6T_{\Lambda}))$ on the $g_{0}=g_{12}$ plane
  is  the boundary of the transition from the thermal  to the condensed phase. 
     This line is part of the critical surface,  illustrated as the transparent green plane in Fig.~\ref{fig:flow-3d}, below which the chemical potential flows to $- \infty$, and above which it flows to $+ \infty$. 
    We note that this surface bends down to negative values of $\mu$ and $g_{0}$, which is a prerequisite for the first order transition for $g_{12}>g_{0}$, discussed below. 
  
%  Thus, the plane spanned by the two phase-boundary lines on the $g_{0}=g_{12}$ and $g_{12}=0$ plane, illustrated as the transparent green plane in Fig.~\ref{fig:flow-3d}, is qualitatively the border of the phase transition.
%%
\begin{figure}
\includegraphics[height=6.6cm]{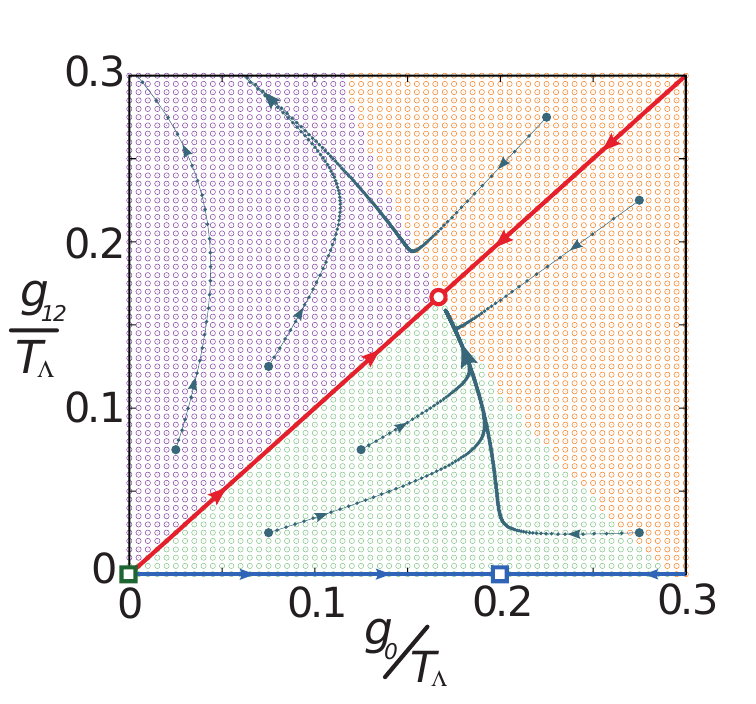}
\caption{The dark blue lines are flow trajectories projected on the $g_0$-$g_{12}$ plane, obtained by numerical integration of (\ref{eqn:flow-final1-alt}-\ref{eqn:flow-final3-alt}). The empty circles and squares are the locations of the three fixed points projected on the plane. The initial values of the trajectories are shown as blue dots. The initial value of $\mu_{\Lambda}$ is $\mu_{\Lambda}(0)=1/3$. $\mu_{\Lambda}$ flows to positive (green and purple circles) values, which indicates a condensed phase, or to negative values (orange circles), indicating a thermal phase. Furthermore, the regime of a discontinuous transition is indicated with purple circles, and the regime of a continuous transition with green circles. }
\label{fig:flow-in-v0-v12-plane}
\end{figure}

Finally, the flow behavior perpendicular to the $g_{0}=g_{12}$ plane shows marginal behavior in the first-order analysis near the third fixed point, see Appendix~\ref{flow-first-order}. 
 We integrate the RG equations to study the RG flow further. 
 In Fig.~\ref{fig:flow-in-v0-v12-plane}, we depict the RG flow in the  $g_0$-$g_{12}$ plane, that is created by the Eqs.~(\ref{eqn:flow-final1-alt}), (\ref{eqn:flow-final2-alt}) and (\ref{eqn:flow-final3-alt}). As an example, we fixed the initial value of the chemical potential to $\mu_{\Lambda}(0)=1/3$.
 We choose several initial values of $(g_0,g_{12})$ to  depict the trajectories, to illustrate the flow.  
 %mapped onto the $g_0$-$g_{12}$ plane, and illustrate $\mu_{\Lambda}(l)$ flowing toward large positive as 

  We also indicate the three fixed points, 
 as well as  whether the chemical potential flows to $+ \infty$, or towards $- \infty$, via color coding:  
 %green or purple circles, and toward negative as orange circles, respectively.
%We choose different initial values of $(g_0,g_{12})$ with fixed initial $\mu_{\Lambda}(0)=1/4$ to numerically integrate Eq.~(\ref{eqn:flow-final1-alt}), (\ref{eqn:flow-final2-alt}) and (\ref{eqn:flow-final3-alt}). 
%When the chemical potential flows to $+ \infty$, it indicates condensation, when it flows to to $- \infty$, it indicates that the system forms a thermal phase. 
% The physical meaning of the chemical potential flowing to $+ \infty$ corresponds to a phase transition to a condensed phase, and flowing to $- \infty$ corresponds to a thermal-gas phase. 
 The orange colored region  corresponds to the $\mu \rightarrow -\infty$ regime, whereas the green and purple regions indicate $\mu \rightarrow + \infty$. 
  The latter regime is further divided into two subregimes, which are distinguished 
  by the asymptotic behavior of $g_{0}$ and $g_{12}$ under the flow. 
  If the flow approaches the fixed point $(g_{0}^{\ast},g_{12}^{\ast})=(\varepsilon/(6T_{\Lambda}),\varepsilon/(6T_{\Lambda}))$ asymptotically, the initial point is colored green. 
  This occurs for $0<g_{12}(0)/g_{0}(0)\leq1$.
    If the flow is repelled by this fixed point, and $g_{0}$ asymptotically flows to a negative value, the initial point is colored purple.    This occurs for $g_{12}(0)/g_{0}(0)>1$.

  For the regime $0<g_{12}(0)/g_{0}(0)\leq1$, the critical behavior is controlled by the fixed point at  $(g_{0}^{\ast},g_{12}^{\ast})=(\varepsilon/(6T_{\Lambda}),\varepsilon/(6T_{\Lambda}))$.
    The flow towards to this point separates into an initial, fast flow onto the marginal surface, connecting the second and third fixed point, on which it flows logarithmically slow, when parametrized by $l$. The chemical potential on the other hand diverges exponentially fast to either $\pm \infty$. When its absolute value reaches unity, the flow moves out of its range of validity, and we terminate it. Therefore, the RG flow, while asymptotically moving towards the Heisenberg fixed point, predicts a physical state for which $g^{*}_{12}<g^{*}_{0}$. This suggests that the system breaks both $U(1)$ symmetries, while the $\mathbb{Z}_{2}$ is preserved. 
        The perturbation  $\sim (g_{0} - g_{12}) (|\phi_{1}|^{2} - |\phi_{2}|^{2})^{2}$ is therefore dangerously irrelevant: It modifies the ground state of the system in a qualitative way from the state that it would have if the magnitude of $g_{0} - g_{12}$ was identically zero. 
    %  Since the effective free energy is described by a quartic Landau-Ginzburg theory, this phase transition of two weakly coupled condensation transitions is continuous. 
%However, according to the analysis in Huang-Yang-Luttinger approach, the system will flow into a critical situation between the $\mathbb{Z}_{2}$ symmetry breaking and preserving in the regime $0<g_{12}(0)/g_{0}(0)\leq1$
  %   The results indicates that the critical behavior in the regime is determined by the two-component Bose gas with $g_{12}=g_{0}$ at low temperature. Since the effective free energy is well-described by a quartic theory, a thermal-gas to a condensate phase transition belongs to a continuous type. However, according to the analysis in Huang-Yang-Luttinger approach, the system will flow into a critical situation between the $\mathbb{Z}_{2}$ symmetry breaking and preserving in the regime $0<g_{12}(0)/g_{0}(0)\leq1$
  The second order phase boundary between this phase with two condensates, each breaking a $U(1)$ symmetry, and the thermal phase, is indicated by the green and orange circles.
  This phase boundary shifts to larger values of $g_0$ and $g_{12}$,   when the initial value of $\mu_{\Lambda}$ is increased.
    When it is reduced, the boundary shifts to smaller values. This indicates that the condensation transition can tolerate larger repulsive interactions, when the chemical potential is larger.

\begin{figure}
\includegraphics[height=6.0cm]{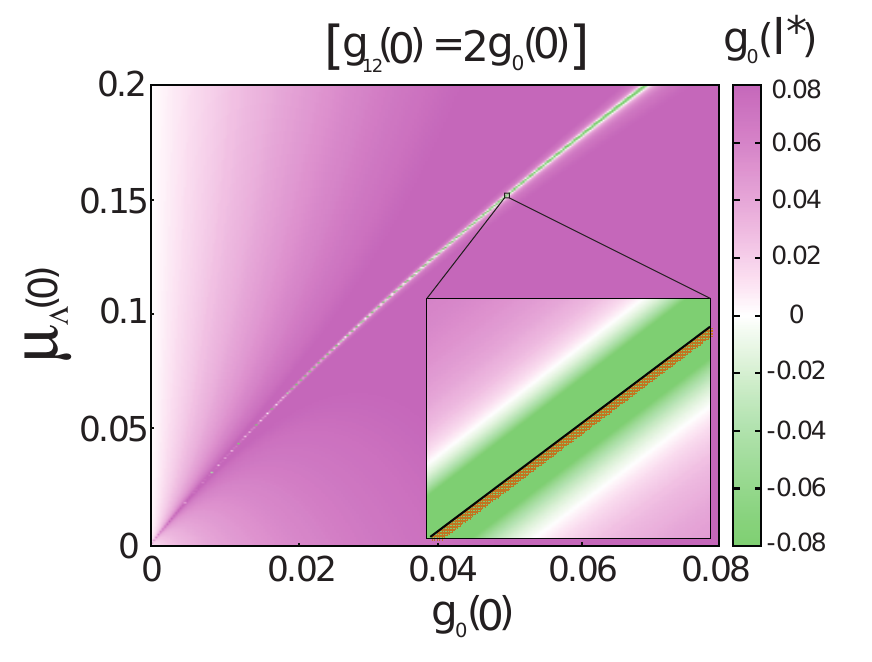}
\caption{We depict $g_{0}(\ell^{*})$ as a function of $g_{0}(0)$ and the chemical potential. Near the critical surface, $g_{0}$ is renormalized to a negative value. In the inset we show the direct vicinity of the critical surface, and also indicate the regime for which $g_{0}^{2}/|\mu|<0.1$, as orange. This regime which exists below the critical surface for any initial value, is responsible for the first order transition of the system.  }
\label{fig:first-order-transition}
\end{figure}
\begin{figure}
\includegraphics[height=6.7cm]{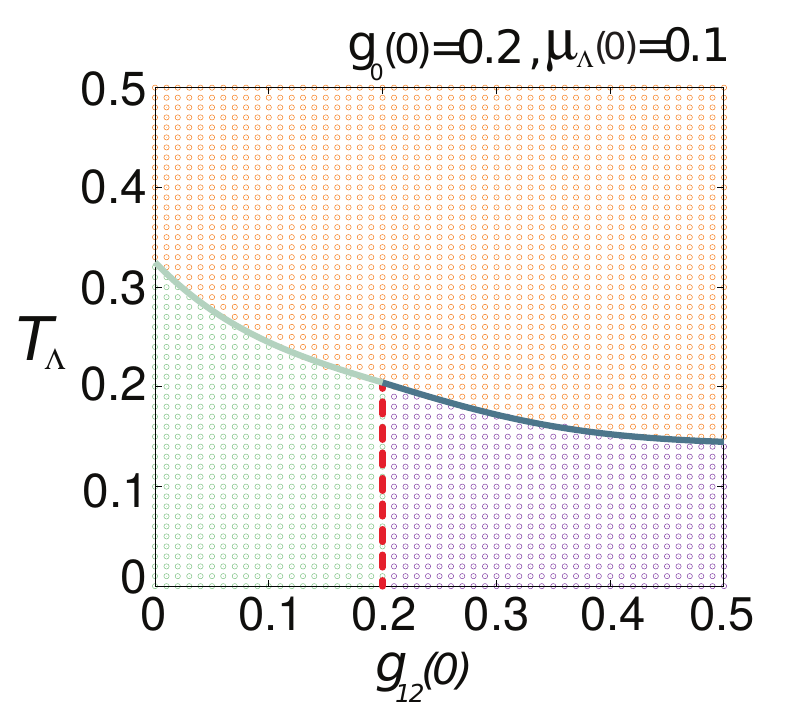}
\caption{Phase diagram of the system  as a function of $g_{12}(0)$ and $T_{\Lambda}$, with fixed $g_{0}(0)=0.2$ and $\mu_{\Lambda}(0)=0.1$. The orange circles denoted the thermal gas phase. The green circles represent the state of two condensates, each breaking a $U(1)$ symmetry, and the purple circles represent a state with a broken $U(1)$ symmetry and a broken chiral, $\mathbb{Z}_2$ symmetry.
 The purple (green) line  represents the discontinuous (continuous) phase transition from the thermal gas phase.}
\label{fig:PD2}
\end{figure}

For $g_{12}(0)/g_{0}(0)>1$, the parameters are initially attracted to a line that is approximately perpendicular to the line $g_{0} = g_{12}$,
 as visible in  Fig.~\ref{fig:flow-in-v0-v12-plane}. Once in the vicinity of this line, the parameters are now repelled by the Heisenberg fixed point.
  In particular, $g_{0}$ is renormalized to a negative value, while $g_{12}$ increases and remains positive.
      This indicates the breakdown of the quartic, effective field theory, as the energy of the system is no longer bounded from below. Higher order terms, such as a $\phi^{6}$ term:
   \bea
   F_{6} &\sim & g_{6} \int d \bR (|\phi_{1}|^{6} + |\phi_{2}|^{6})
   \eea
   have to be included to provide a stable description of the system.
    Furthermore, it indicates that a first-order phase transition can occur, Ref.  \cite{Domany}. 
%WM
We note that this $\phi^6$ term is an effective three-body interaction. It is generated by integrating out high-energy excitations, and results from virtual two-body collisions. For a deep optical lattice, this has been discussed in Refs.~\cite{Johnson,Will}. This scenario is demonstrated in Fig. \ref{fig:first-order-transition}. For a first order transition to occur in this system, an effective free energy with three distinct minima has to emerge, similar to the example in Fig. \ref{fig:MFA} (c ). For this to occur, the chemical potential has to flow to a negative value, as demonstrated above, $g_{0}$ has to flow to a negative value, as shown above, and finally the magnitude of $g_{0}$, $\mu$ and $g_{6}$ has to be such that three minima occur, and the side minima emerge as the global minima. As can be checked, for this $g_{0}^{2}/|\mu| \gtrsim g_{6}$ is required.
  
In Fig. \ref{fig:first-order-transition} we show the $\mu$-$g_{0}$ plane, where we set $g_{12} = 2 g_{0}$, which  is the plane relevant to the triangular lattice system. We depict the magnitude of $g_{0}$ for $\ell^{*}$, for which $|\mu(\ell^{*})|=1$. Near the critical surface, $g_{0}$ is negative for any initial value.
 Furthermore, we depict the regime for which $g_{0}^{2}/|\mu| \gtrsim 0.1$. This, and any other choice for $g_{6}$, results in a narrow regime below the entire critical surface. 
  It is this scenario that suggests a first order transition. We note that a bosonic system of constant density has a monotonously increasing chemical potential, as the temperature is lowered. When the chemical potential reaches the critical surface, the system condenses.
   For the system of two bosonic fields, and for $g_{12} >g_{0}$, however, before the chemical reaches the critical surface, the effective action develops side minima, which results in a first order transition.

%
%we find that $g_{0}$ and $g_{12}$ tend to flow close to the fixed point at $g_{0}^{\ast}=g_{12}^{\ast}$ in the beginning, but eventually $g_{0}$ flows to a negative value, while $g_{12}$ grows and remains positive. The flow to negative values of $g_0$ causes free energy unbounding. It indicates the failure of the effective quartic theory (\ref{hie}). Higher-order terms in the field theory decomposition or a three-body interaction in the original Hamiltonian should be considered. Moreover, an unbounding of the free energy in a quartic theory also causes the occurrence of a discontinuous phase transition \cite{Domany}. Thus, we conclude in the regime $g_{12}(0)/g_{0}(0)>1$, the phase transition from thermal gases to condensate is yielded to be discontinuous accompanied with the $\mathbb{Z}_{2}$ symmetry breaking.

%DISCUSS $\phi^{6}$ TERM, SEE FIG. \ref{fig:first-order-transition}

%
%Finally, two special cases are indicated by the straight red and a blue lines  in Fig.~\ref{fig:flow-in-v0-v12-plane}. The blue line describes the flow

%The projected flows, illustrated as a straight red line and a blue line in Fig.~\ref{fig:flow-in-v0-v12-plane}, reflect our analysis in the vicinity of fixed points: when the initial values satisfy $g_{0}(0)=g_{12}(0)$ and $g_{12}(0)=0$, the RG flow remains on the $g_{0}=g_{12}$ and $g_{12}=0$ plane, respectively. 

\subsection{Phase diagram}
\begin{figure}
\includegraphics[height=8.cm]{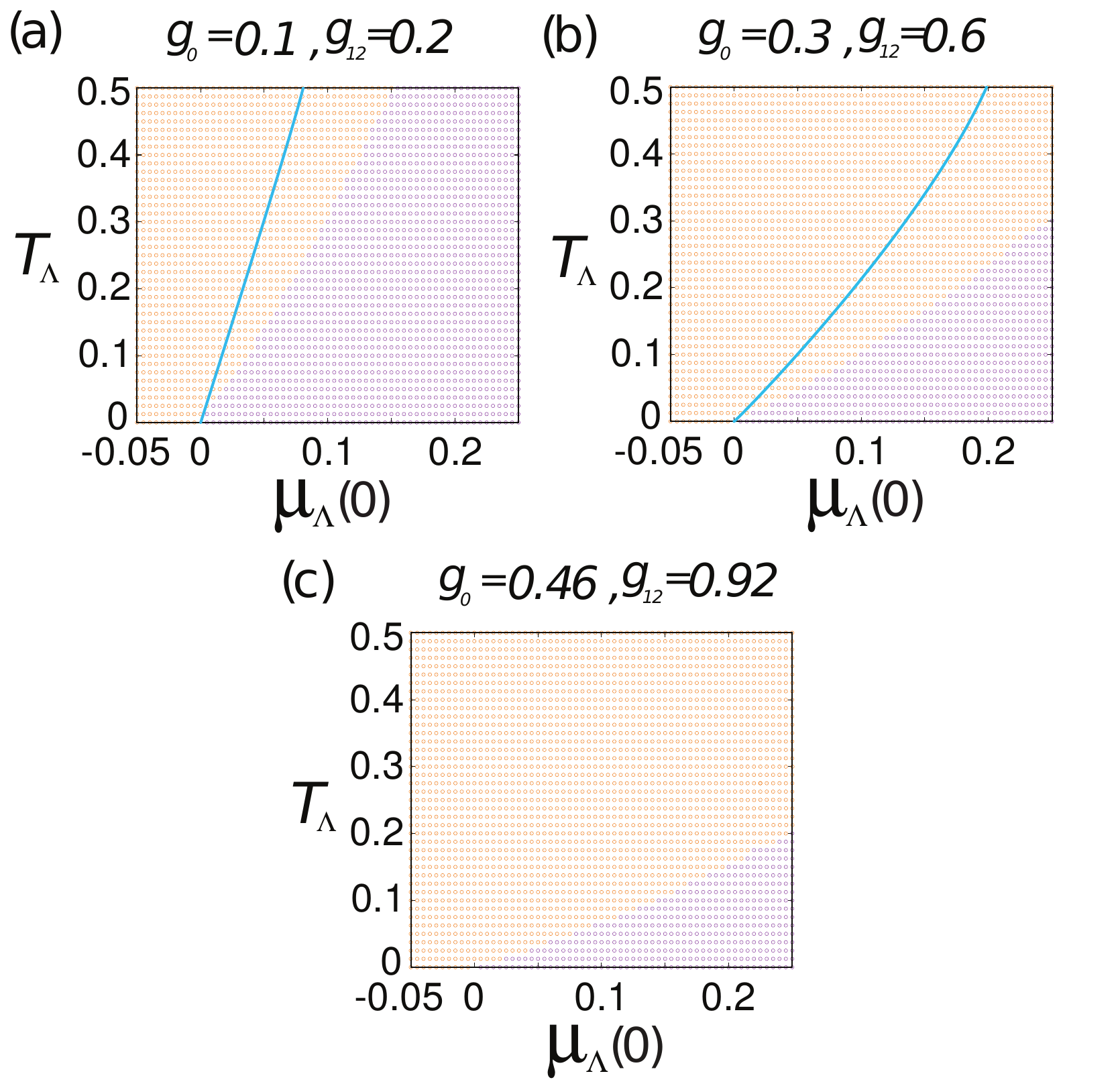}
\caption{Phase diagrams as a function of  $T_{\Lambda}$ and $\mu_{\Lambda}(0)$, with (a) $g_0(0)=0.1$, $g_{12}(0)=0.2$, (b)$g_0(0)=0.3$, $g_{12}(0)=0.6$ and (c)$g_0(0)=0.46$, $g_{12}(0)=0.92$ (the interaction found in the experiment~\cite{Struck13}). Orange circles again label the thermal phase, the condensed phase is depicted with purple circles. 
 To illustrate the renormalization of the critical temperature by $g_{12}$, we 
 depict the blue line, which represents the phase boundary for (a)$g_0(0)=0.1$, $g_{12}(0)=0$ and (b) $g_{0}(0)=0.3$, $g_{12}(0)=0$.}
\label{fig:PD3}
\end{figure}

In Figs. \ref{fig:PD2} and \ref{fig:PD3} we show the resulting phase diagram of the system. 
 In Fig. \ref{fig:PD2} we keep the chemical potential and $g_{0}$ fixed, and vary $g_{12}$ and the temperature. 
    For high temperatures, the system is in the thermal gas phase, of any value of $g_{12}$. This regime is labelled as orange circles. 
  As the temperature is lowered, the system condenses, either into a phase of two condensates, each breaking a $U(1)$ symmetry, or into a chiral condensate phase, which breaks a $U(1)$ and the $\mathbb{Z}_2$ symmetry. 
This transition occurs at $g_{12}(0)=g_{0}(0)$, as illustrated as the red dashed line.
Furthermore, the order of the phase transition changes, as $g_{12}$ is increased. For $g_{12}(0)<g_{0}(0)$ the system undergoes a second order phase transition, for $g_{12}(0)>g_{0}(0)$ a first order transition. 
 We also note that the critical temperature is reduced with increasing $g_{12}(0)$. 
 To illustrate the influence of the inter-component interactions $g_{12}(0)$ on the critical temperature further, we show the phase diagram in the $T_{\Lambda}$-$\mu_{\Lambda}(0)$ plane in Fig.~\ref{fig:PD3}. We note that the phase diagram for the interaction found in the experiment~\cite{Struck13} is illustrated versus temperature and chemical potential in Fig.~\ref{fig:PD3}(c).
 
% For $g_0(0)=0.1$ and $g_0(0)=0.3$ with both $g_{12}(0)=0$, the phase boundaries are reminisced to single-component $\phi^4$-theory, and illustrated as the blue line in the figures. However, while we turn on $g_{12}(0)$ the phase boundaries move to lower temperatures and larger chemical potentials. 

% 
% For large $T_{\Lambda}$, a regime of a thermal-gas phase is labelled as orange circles. By lowering temperature, we denote a condensate phase with continuous and discontinuous phase transition from thermal gases as green and purple circles respectively. As mentioned above, the boundary of a continuous and discontinuous phase transition is at $g_{12}(0)=g_{0}(0)$, as illustrated as the red dashed line. 

%With a fixed $g_{0}(0)=0.2$ and $\mu_{\Lambda}(0)=0.1$, we plot the phase diagram on the $g_{12}(0)$-$T_{\Lambda}$ plane in Fig.~\ref{fig:PD3}. 

%%
%%

%%%%%%%%%%%%%%%%%%%%%%%%
%%   Summary        %%%%%%%%%%

\section{Conclusions}\label{conc}
%
%1. relative systems in XY model 
%
In this article, we have studied the critical behavior of Bose-Einstein condensation in a frustrated, triangular lattice. 
Using a renormalization group approach, we have demonstrated that the phase transition is of first order. At this transition, the system breaks both a $U(1)$ symmetry and a $\mathbb{Z}_2$, simultaneously, resulting in a condensation with chiral order.
We achieve this insight by mapping the system onto a complex $\phi^{4}$ theory consisting of two complex fields. This field theory has a  $U(1) \times U(1) \times \mathbb{Z}_2$ symmetry, corresponding to the $U(1)$ symmetry of each complex field, as well as a $\mathbb{Z}_2$ symmetry of exchanging the two fields.
We demonstrate that the critical behavior of this effective field theory is controlled by the ratio of the  inter-component $V_{12}$ and the intra-component $V_{0}$ interaction strength.  For $V_{12}<V_{0}$ the system undergoes a second order phase transition, in which both $U(1)$ symmetries are broken, while preserving the $\mathbb{Z}_{2}$ symmetry.   For $V_{12}>V_{0}$ the system undergoes a first transition, in which one $U(1)$ symmetry and one $\mathbb{Z}_{2}$ symmetry are broken.
 The latter scenario is reflected in the renormalization group as follows: Rather than being attracted by a fixed point of the flow, which controls the critical behavior, the parameters of the system are repelled by a fixed point, and the emergent free energy becomes unbounded from below. By considering the properties of this flow, this instability can be identified with a first order transition, because the parameters are renormalized in such a way that additional side minima can appear, as in a $\phi^{6}$ theory, and eventually become the global minima. Experimentally, the first order character of the transition could be observed by measuring the condensate fraction and the chiral magnetization as a function of temperature. 
 This intriguing critical behavior, in which condensation occurs as a first order transition, could also occur in numerous other cold atom systems, in which the dispersion of the system has more than one minimum. Our study would apply in an analogous manner, and would therefore be of crucial guidance to understand  the critical behavior in such systems. In the end, we note that our analysis might be motivated primarily by the Sengstock experiment, but that this type of analysis similarly applies to other frustraed systems~\cite{Wirth11}.

%%%%%%%%%%%%%%%%%%%%%%%%
%%   Acknowledgments        %%%%%%%%%%
\begin{acknowledgments}
We acknowledge support from the Deutsche Forschungsgemeinschaft through
the SFB 925 and the Hamburg Centre for Ultrafast Imaging, and from the Landesexzellenzinitiative Hamburg, which is supported by the Joachim Herz Stiftung.  
 WMH acknowledges support from the Ministry of Science and Technology in Taiwan through grant MOST104-2112-M-005-006-MY3. 

\end{acknowledgments}
%
%%%%%%%%%%%%%%%%%%%%%%%%%%%%%

\newpage

%%%%%%%%%%%%%%%%%%%%%%%%
%%   appendix              %%%%%%%%%%
%%%%%%%%%%%%%%%%%%

\appendix

%%%%%%%%%%%%%%%%%%%
%    Appendix 1    %%%%%%%%%%%%%%%%%%
\section{Derivation of the flow equations}\label{RGderivation}

\begin{figure}[b!]
\includegraphics[height=2cm]{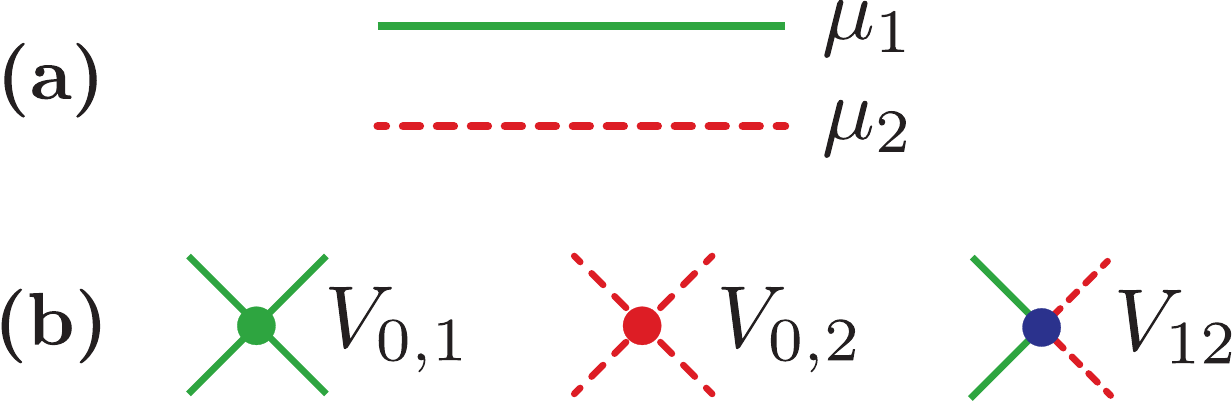}
\caption{(a) we illustrate the propagators of the first and second components as the green solid and red dashed lines respectively, and (b) corresponding elementary vertices of the theory.}
\label{fig:feynman-diagrams}
\end{figure}

In this Appendix, we derive the renormalization group equations for the free energy functional of interest (\ref{eqn:free-energy-0},\ref{eqn:free-energy-I}). For reasons of clarity, we rewrite the free energy $F_{L}[\phi^{*}, \phi]=F_{L,0}+F_{L,I}$ according to:
\begin{align}
F_{L,0} & = \int d^3 \bm{R} \sum_{j=1,2}\Bigg[ \frac{\hbar^{2}}{2m^*}\left| \nabla \phi_{j} \right|^{2} -\mu_{j} \left| \phi_{j} \right|^{2} \Bigg], \\
F_{L,I} & = \int d^{3} \textbf{R} \left( \frac{V_{0,1}}{2} \left| \phi_{1} \right|^{4} + \frac{V_{0,2}}{2} \left| \phi_{2} \right|^{4} + V_{12} \left| \phi_{1} \right|^{2} \left| \phi_{2} \right|^{2} \right) .
\end{align}
This will make it easier to identify distinct terms in the perturbative expansion later on. At the end of the calculation, we will set $\mu_{1}=\mu_{2}=\mu$ and $V_{0,1}=V_{0,2}=V_{0}$. Pictorially, we can represent the two order parameters $\phi_{1}$ and $\phi_{2}$ as colored lines, where the propagator corresponds to the respective chemical potentials and the coupling constants appear as the elementary vertices of the theory (Fig. \ref{fig:feynman-diagrams}).
Assuming the interactions to be weak, the corrections to the propagators and vertices can be obtained using the cumulant expansion of the renormalized free energy functional \cite{Stoof09,Wilson}.

The first-order correction is given by
\begin{align}
\left< F_{L,I} \right>_{0,>} ,
\label{eqn:first-order-correction}
\end{align}
where the expectation value is defined as
\begin{align}
\nonumber
& \left< \cdots \right>_{0,>} \equiv \frac{1}{Z_{1,>} Z_{2,>}} \times \\
\nonumber
& \int \int d \left[ \phi_{1,>}^{\ast} \right] d \left[ \phi_{1,>} \right] d \left[ \phi_{2,>}^{\ast} \right] d \left[ \phi_{2,>} \right] \left( \cdots \right) \times \\
\nonumber
& \hspace{10mm} e^{- \beta \int d^{3} \textbf{R} \left( \frac{\hbar^{2}}{2m_{1}} \left| \nabla \phi_{1,>} \right|^{2} - \mu_{1} \left| \phi_{1,>} \right|^{2} \right)} \times \\
& \hspace{10mm} e^{- \beta \int d^{3} \textbf{R} \left( \frac{\hbar^{2}}{2m_{2}} \left| \nabla \phi_{2,>} \right|^{2} - \mu_{2} \left| \phi_{2,>} \right|^{2} \right)} ,
\label{eqn:expectation-value}
\end{align}
with normalization constants $Z_{1,>}$ and $Z_{2,>}$.

To compute the first-order corrections, we split the two complex order parameters into a high-energy and a low-energy part
\begin{align}
\phi_{i} = \phi_{i,<} + \phi_{i,>} \quad \text{with $i=1,2$} ,
\end{align}
where:
\begin{align}
& \phi_{i,<} = \sum_{k<\Lambda/b} \phi_{i, \textbf{k}} \frac{e^{i \textbf{k} \textbf{R}}}{\sqrt{\mathcal{V}}} \quad \text{with $i=1,2$} , \\
& \phi_{i,>} = \sum_{\Lambda/b<k<\Lambda} \phi_{i, \textbf{k}} \frac{e^{i \textbf{k} \textbf{R}}}{\sqrt{\mathcal{V}}} \quad \text{with $i=1,2$} .
\end{align}
Here, $\mathcal{V}$ is the volume of the system, $\Lambda$ is the high-energy cutoff of the theory and $b$ is some number bigger than one, such that the new cutoff after one RG step is lowered compared to the old one. $\Lambda_{b} = \Lambda/b$ is also referred to as the running cutoff. As can be seen from (\ref{eqn:expectation-value}), the Fourier components of the order parameters that correspond to high-energy excitations are integrated out to obtain corrections for the renormalized free energy functional.

\begin{figure}[t!]
\includegraphics[height=2cm]{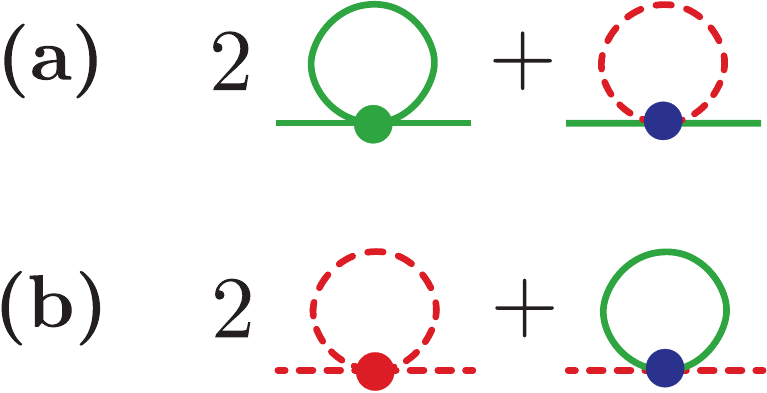}
\caption{The first-order corrections of the RG method to (a) $\mu_{1}$ and (b) $\mu_{2}$.}
\label{fig:mu-correction}
\end{figure}

For each of the three addends in (\ref{eqn:first-order-correction}), the splitting of the order parameter results in $16$ terms. Most terms vanish because of the Gaussian integration in (\ref{eqn:expectation-value}), since they involve an odd number of high-energy parts $\phi_{i,>}$ of the order parameter. Also, $\phi_{i,>}$ must always appear in combination with its complex conjugate to yield a non-zero contribution. Of the remaining terms, only those are of interest that contribute corrections to the chemical potentials (in contrast to just shifting the renormalized free energy functional by a constant factor). Therefore, the only relevant terms for the flow equations are:
\begin{align}
4 \frac{V_{0,1}}{2} & \int d^{3} \textbf{R} \, \phi_{1,<} \phi_{1,<}^{\ast} \left< \phi_{1,>}^{\ast} \phi_{1,>} \right>_{0,>} , \label{eqn:first-order-one} \\
4 \frac{V_{0,2}}{2} & \int d^{3} \textbf{R} \, \phi_{2,<} \phi_{2,<}^{\ast} \left< \phi_{2,>}^{\ast} \phi_{2,>} \right>_{0,>} , \label{eqn:first-order-two} \\
V_{12} & \int d^{3} \textbf{R} \, \phi_{2,<} \phi_{2,<}^{\ast} \left< \phi_{1,>}^{\ast} \phi_{1,>} \right>_{0,>} , \label{eqn:first-order-three} \\
V_{12} & \int d^{3} \textbf{R} \, \phi_{1,<} \phi_{1,<}^{\ast} \left< \phi_{2,>}^{\ast} \phi_{2,>} \right>_{0,>} . \label{eqn:first-order-four}
\end{align}
We can see that the $V_{12}$-interaction mixes in corrections from $\mu_{1}$ to $\mu_{2}$ and vice versa. The remaining evaluation of the expectation value can be carried out in Fourier space straightforwardly
\begin{align}
\nonumber
\left< \phi_{i,>}^{\ast} \phi_{i,>} \right>_{0,>} & = \frac{1}{\mathcal{V}} \sum_{\Lambda/b<k<\Lambda} \frac{1}{\beta \left( \varepsilon_{i,\textbf{k}} - \mu_{i} \right)} \\
& = \int_{\Lambda/b}^{\Lambda} \frac{d \textbf{k}}{(2 \pi)^{d}} \frac{k_{B} T}{\varepsilon_{i,\textbf{k}} - \mu_{i}} ,
\end{align}
where, in the last step, the continuum limit of the Fourier sum is taken in $d$ dimensions and $\varepsilon_{i,\textbf{k}} = \hbar^{2} \textbf{k}^{2} / (2 m_{i})$. To mimic our original system, a rescaling of the Fourier components $k \rightarrow k/b$ must be performed after one RG step. The shell in Fourier space that is integrated out is taken to be infinitesimally thin with thickness $d \Lambda = \Lambda dl$ and area $2 \pi^{d/2} \Lambda^{d-1} / \Gamma \left( \frac{d}{2} \right)$, where $l=\ln(b)$ and $\Gamma \left( \cdots \right)$ is the Gamma function. With this, the change in $\mu_{1}$ and $\mu_{2}$ after one RG step caused by the corrections (\ref{eqn:first-order-one}-\ref{eqn:first-order-four}) can be expressed as
\begin{align}
\nonumber
d \mu_{i} = & - 2 V_{0,i} \frac{\Lambda^{d}}{(2 \pi)^{d}} \frac{2 \pi^{d/2}}{\Gamma \left( \frac{d}{2} \right)} \frac{k_{B} T}{e^{-2l} \varepsilon_{i,\Lambda} - \mu_{i}} e^{-ld} dl \\
& - V_{12} \frac{\Lambda^{d}}{(2 \pi)^{d}} \frac{2 \pi^{d/2}}{\Gamma \left( \frac{d}{2} \right)} \frac{k_{B} T}{e^{-2l} \varepsilon_{\bar{i},\Lambda} - \mu_{\bar{i}}} e^{-ld} dl ,
\end{align}
where, as before, $i$ is either one or two and $\bar{i}$ is the negation of $i$. These first-order corrections are depicted in Fig. \ref{fig:mu-correction}.
To obtain the desired flow equations, we have to rescale the chemical potentials and the coupling constants according to:
\begin{align}
\nonumber
\mu_{i} & \rightarrow \mu_{i} e^{-2l} , \\
\nonumber
V_{0,i} & \rightarrow V_{0,i} e^{-(4-d)l} , \\
V_{12} & \rightarrow V_{12} e^{-(4-d)l} .
\label{eqn:rescale}
\end{align}
The flow equations for $\mu_{1}$ and $\mu_{2}$ take the form:
\begin{align}
\nonumber
\dfrac{d \mu_{i}}{dl} = 2 \mu_{i} & - 2 V_{0,i} \frac{\Lambda^{d}}{(2 \pi)^{d}} \frac{2 \pi^{\frac{d}{2}}}{\Gamma \left( \frac{d}{2} \right)} \frac{k_{B} T}{\varepsilon_{i,\Lambda} - \mu_{i}} \\
& - V_{12} \frac{\Lambda^{d}}{(2 \pi)^{d}} \frac{2 \pi^{\frac{d}{2}}}{\Gamma \left( \frac{d}{2} \right)} \frac{k_{B} T}{\varepsilon_{\bar{i},\Lambda} - \mu_{\bar{i}}} .
\label{eqn:flow1}
\end{align}
Setting $\mu_{1}=\mu_{2}=\mu$ and $V_{0,1}=V_{0,2}=V_{0}$ results in the RG equation for the chemical potential.

Now we turn to the second-order correction within the cumulant expansion, which is given by:
\begin{align}
\frac{1}{2} & \left( \left< \left( F_{L,I} \right)^{2} \right>_{0,>} - \left< F_{L,I} \right>_{0,>}^{2} \right) .
\label{eqn:second-order-correction}
\end{align}
As usual, corrections to $V_{0,1}$, $V_{0,2}$ and $V_{12}$ will take the form of connected diagrams. Disconnected diagrams are killed off by the second addend in (\ref{eqn:second-order-correction}). We can group the resulting terms into those being proportional to either $V_{0,1}^{2}$, $V_{0,2}^{2}$, $V_{12}^{2}$, $V_{0,1} V_{0,2}$, $V_{0,1} V_{12}$ or $V_{0,2} V_{12}$. In principle, for each of these six groups, $256$ terms must be considered. As in the calculation for the first-order correction, the analysis simplifies, since terms with an odd number of high-energy parts $\phi_{i,>}$ of the order parameter vanish. Also, corrections to the coupling constants should have two incoming and two outgoing low-energy parts $\phi_{i,<}$. Terms with one or three incoming or outgoing $\phi_{i,<}$ factors do not contribute.

Some remaining non-vanishing terms require additional discussion. Terms with only two $\phi_{i,<}$ factors would yield further corrections to the chemical potentials. These corrections are of higher order in the coupling constants than the first-order corrections already considered and therefore can be neglected. Terms with six $\phi_{i,<}$ factors imply an interaction proportional to $\phi^{6}$ not present in the original free energy functional. These terms are usually disregarded in the framework of the $\varepsilon$-expansion \cite{Wilson}. As before, constant shifts to the free energy functional are not of interest.

Strictly, the $\varepsilon$-expansion is only valid near four dimensions. Nevertheless it proved to be successful in lower dimensions as well, especially in the important three-dimensional case. Therefore we are going to use the results obtained using the $\varepsilon$-expansion for a three-dimensional system in the main text.

The terms that are relevant for the flow equations will contain an average taken over four fields:
\begin{align}
\left< \phi_{i,>}^{\ast} \phi_{i,>} \phi_{j,>}^{\ast} \phi_{j,>} \right>_{0,>} .
\end{align}
Because of the Gaussian integration, this expression is split into products of averages of two fields. When $i=j$, there are two possibilities to contract the fields and a factor of $2$ must be introduced.

Now we collect the interesting terms group by group. The two groups proportional to $V_{0,i}^{2}$ contribute the following connected diagrams:
\begin{align}
\nonumber
16 \frac{V_{0,i}^{2}}{4} & \int \int d^{3} \textbf{R} \, d^{3} \textbf{R}^{\prime} \, \phi_{i,<}(\textbf{R}) \phi_{i,<}^{\ast}(\textbf{R}) \phi_{i,<}(\textbf{R}^{\prime}) \phi_{i,<}^{\ast}(\textbf{R}^{\prime}) \times \\
& \left< \phi_{i,>}^{\ast}(\textbf{R}) \phi_{i,>}(\textbf{R}^{\prime}) \right>_{0,>} \left< \phi_{i,>}^{\ast}(\textbf{R}^{\prime}) \phi_{i,>}(\textbf{R}) \right>_{0,>} , \label{eqn:second-order-v-one} \\
\nonumber
4 \frac{V_{0,i}^{2}}{4} & \int \int d^{3} \textbf{R} \, d^{3} \textbf{R}^{\prime} \, \phi_{i,<}(\textbf{R}) \phi_{i,<}^{\ast}(\textbf{R}^{\prime}) \phi_{i,<}(\textbf{R}) \phi_{i,<}^{\ast}(\textbf{R}^{\prime}) \times \\
& \left< \phi_{i,>}^{\ast}(\textbf{R}) \phi_{i,>}(\textbf{R}^{\prime}) \right>_{0,>} \left< \phi_{i,>}^{\ast}(\textbf{R}) \phi_{i,>}(\textbf{R}^{\prime}) \right>_{0,>} . \label{eqn:second-order-v-two}
\end{align}

The group of terms proportional to $V_{12}^{2}$ results in a variety of connected diagrams:
\begin{align}
\nonumber
V_{12}^{2} & \int \int d^{3} \textbf{R} \, d^{3} \textbf{R}^{\prime} \, \phi_{2,<}(\textbf{R}) \phi_{2,<}^{\ast}(\textbf{R}) \phi_{2,<}(\textbf{R}^{\prime}) \phi_{2,<}^{\ast}(\textbf{R}^{\prime}) \times \\
& \left< \phi_{1,>}^{\ast}(\textbf{R}) \phi_{1,>}(\textbf{R}^{\prime}) \right>_{0,>} \left< \phi_{1,>}^{\ast}(\textbf{R}^{\prime}) \phi_{1,>}(\textbf{R}) \right>_{0,>} , \label{eqn:second-order-v12-one} \\
\nonumber
V_{12}^{2} & \int \int d^{3} \textbf{R} \, d^{3} \textbf{R}^{\prime} \, \phi_{1,<}(\textbf{R}) \phi_{1,<}^{\ast}(\textbf{R}) \phi_{1,<}(\textbf{R}^{\prime}) \phi_{1,<}^{\ast}(\textbf{R}^{\prime})\times  \\
& \left< \phi_{2,>}^{\ast}(\textbf{R}) \phi_{2,>}(\textbf{R}^{\prime}) \right>_{0,>} \left< \phi_{2,>}^{\ast}(\textbf{R}^{\prime}) \phi_{2,>}(\textbf{R}) \right>_{0,>} , \label{eqn:second-order-v12-two} \\
\nonumber
2 V_{12}^{2} & \int \int d^{3} \textbf{R} \, d^{3} \textbf{R}^{\prime} \, \phi_{1,<}(\textbf{R}^{\prime}) \phi_{1,<}^{\ast}(\textbf{R}) \phi_{2,<}(\textbf{R}^{\prime}) \phi_{2,<}^{\ast}(\textbf{R}) \times \\
& \left< \phi_{1,>}^{\ast}(\textbf{R}^{\prime}) \phi_{1,>}(\textbf{R}) \right>_{0,>} \left< \phi_{2,>}^{\ast}(\textbf{R}^{\prime}) \phi_{2,>}(\textbf{R}) \right>_{0,>} , \label{eqn:second-order-v12-three} \\
\nonumber
2 V_{12}^{2} & \int \int d^{3} \textbf{R} \, d^{3} \textbf{R}^{\prime} \, \phi_{1,<}(\textbf{R}^{\prime}) \phi_{1,<}^{\ast}(\textbf{R}) \phi_{2,<}(\textbf{R}) \phi_{2,<}^{\ast}(\textbf{R}^{\prime}) \times \\
& \left< \phi_{1,>}^{\ast}(\textbf{R}^{\prime}) \phi_{1,>}(\textbf{R}) \right>_{0,>} \left< \phi_{2,>}^{\ast}(\textbf{R}) \phi_{2,>}(\textbf{R}^{\prime}) \right>_{0,>} . \label{eqn:second-order-v12-four}
\end{align}
Analogous to the first-order calculation, the $V_{12}$-interaction mixes in corrections from $V_{0,1}$ to $V_{0,2}$ and vice versa (\ref{eqn:second-order-v12-one},\ref{eqn:second-order-v12-two}). The group of terms proportional to $V_{0,1} V_{0,2}$ does not contain any connected diagrams and therefore does not contribute to the flow equations. This can be easily seen, since a nonzero mixing interaction $V_{12}$ is needed for the two Bose gases to have any influence on one another.

The last two groups of terms we have to consider are those proportional to $V_{0,i} V_{12}$. The connected diagrams are:
\begin{align}
\nonumber
4 \frac{V_{0,i} V_{12}}{2} & \int \int d^{3} \textbf{R} \, d^{3} \textbf{R}^{\prime} \, \phi_{i,<}(\textbf{R}) \phi_{i,<}^{\ast}(\textbf{R}) \phi_{\bar{i},<}(\textbf{R}^{\prime}) \phi_{\bar{i},<}^{\ast}(\textbf{R}^{\prime}) \times \\
& \left< \phi_{i,>}^{\ast}(\textbf{R}) \phi_{i,>}(\textbf{R}^{\prime}) \right>_{0,>} \left< \phi_{i,>}^{\ast}(\textbf{R}^{\prime}) \phi_{i,>}(\textbf{R}) \right>_{0,>} . \label{eqn:second-order-v-v12}
\end{align}
These corrections to $V_{12}$ appear twice in the cumulant expansion (\ref{eqn:second-order-correction}).

\begin{figure}[t!]
\includegraphics[height=3cm]{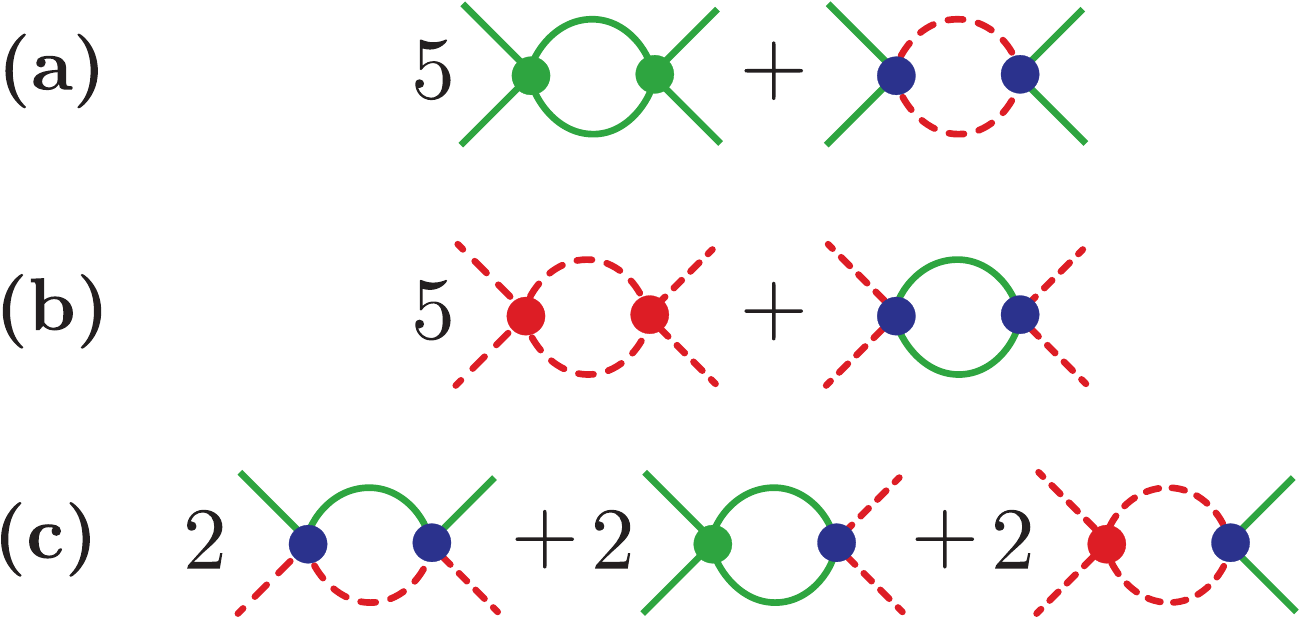}
\caption{The second-order corrections of the RG approach to (a) $V_{0,1}$, (b) $V_{0,2}$ and $V_{12}$.}
\label{fig:v-correction}
\end{figure}

One last complication arises in evaluating the corrections (\ref{eqn:second-order-v-one}-\ref{eqn:second-order-v-v12}), since these corrections include non-local interactions not present in the original free energy functional. Assuming the interaction to be short-ranged, these non-local terms can be neglected within the $\varepsilon$-expansion \cite{Wilson}. Then, the second-order corrections take the form:
\begin{align}
\nonumber
& \left< \phi_{i,>}^{\ast}(\textbf{R}) \phi_{i,>}(\textbf{R}^{\prime}) \right>_{0,>} \left< \phi_{j,>}^{\ast}(\textbf{R}^{\prime}) \phi_{j,>}(\textbf{R}) \right>_{0,>} \\
\simeq & \int_{\Lambda/b}^{\Lambda} \frac{d \textbf{k}}{(2 \pi)^{d}} \left( \frac{k_{B} T}{\varepsilon_{i,\textbf{k}} - \mu_{i}} \right) \left( \frac{k_{B} T}{\varepsilon_{j,\textbf{k}} - \mu_{j}} \right) .
\end{align}
Now we are in the position to quantify the change in $V_{0,1}$, $V_{0,2}$ and $V_{12}$ after one RG step. As before, we rescale the Fourier components and integrate out infinitesimally thin shells in Fourier space:
\begin{align}
\nonumber
d V_{0,i} = & - 5 V_{0,i}^{2} \frac{\Lambda^{d}}{(2 \pi)^{d}} \frac{2 \pi^{\frac{d}{2}}}{\Gamma \left( \frac{d}{2} \right)} \frac{k_{B} T \cdot e^{-ld} dl}{ \left( e^{-2l} \varepsilon_{i,\Lambda} - \mu_{i} \right)^{2}} \\
& - V_{12}^{2} \frac{\Lambda^{d}}{(2 \pi)^{d}} \frac{2 \pi^{\frac{d}{2}}}{\Gamma \left( \frac{d}{2} \right)} \frac{k_{B} T \cdot e^{-ld} dl}{ \left( e^{-2l} \varepsilon_{\bar{i},\Lambda} - \mu_{\bar{i}} \right)^{2}} , \\
\nonumber
d V_{12} = & - 2 V_{12}^{2} \frac{\Lambda^{d}}{(2 \pi)^{d}} \frac{2 \pi^{\frac{d}{2}}}{\Gamma \left( \frac{d}{2} \right)} \frac{k_{B} T \cdot e^{-ld} dl}{\left( e^{-2l} \varepsilon_{1,\Lambda} - \mu_{1} \right) \left( e^{-2l} \varepsilon_{2,\Lambda} - \mu_{2} \right)} \\
\nonumber
& - 2 V_{0,1} V_{12} \frac{\Lambda^{d}}{(2 \pi)^{d}} \frac{2 \pi^{\frac{d}{2}}}{\Gamma \left( \frac{d}{2} \right)} \frac{k_{B} T \cdot e^{-ld} dl}{\left( e^{-2l} \varepsilon_{1,\Lambda} - \mu_{1} \right)^{2}} \\
& - 2 V_{0,2} V_{12} \frac{\Lambda^{d}}{(2 \pi)^{d}} \frac{2 \pi^{\frac{d}{2}}}{\Gamma \left( \frac{d}{2} \right)} \frac{k_{B} T \cdot e^{-ld} dl}{\left( e^{-2l} \varepsilon_{2,\Lambda} - \mu_{2} \right)^{2}} .
\end{align}
These second-order corrections are depicted in Fig. \ref{fig:v-correction}.

Finally, to obtain the flow equations, we have to rescale the chemical potentials and the coupling constants according to (\ref{eqn:rescale}):
\begin{align}
\nonumber
\dfrac{d V_{0,i}}{d l} = & \, (4 - d) V_{0,i} - 5 V_{0,i}^{2} \frac{\Lambda^{d}}{(2 \pi)^{d}} \frac{2 \pi^{\frac{d}{2}}}{\Gamma \left( \frac{d}{2} \right)} \frac{k_{B} T}{ \left( \varepsilon_{i,\Lambda} - \mu_{i} \right)^{2}} \\
& - V_{12}^{2} \frac{\Lambda^{d}}{(2 \pi)^{d}} \frac{2 \pi^{\frac{d}{2}}}{\Gamma \left( \frac{d}{2} \right)} \frac{k_{B} T }{ \left( \varepsilon_{\bar{i},\Lambda} - \mu_{\bar{i}} \right)^{2}} ,
\label{eqn:flow2} \\
\nonumber
\dfrac{d V_{12}}{d l} = & \, (4 - d) V_{12} \\
\nonumber
& - 2 V_{12}^{2} \frac{\Lambda^{d}}{(2 \pi)^{d}} \frac{2 \pi^{\frac{d}{2}}}{\Gamma \left( \frac{d}{2} \right)} \frac{k_{B} T}{\left( \varepsilon_{1,\Lambda} - \mu_{1} \right) \left( \varepsilon_{2,\Lambda} - \mu_{2} \right)} \\
\nonumber
& - 2 V_{0,1} V_{12} \frac{\Lambda^{d}}{(2 \pi)^{d}} \frac{2 \pi^{\frac{d}{2}}}{\Gamma \left( \frac{d}{2} \right)} \frac{k_{B} T}{\left( \varepsilon_{1,\Lambda} - \mu_{1} \right)^{2}} \\
& - 2 V_{0,2} V_{12} \frac{\Lambda^{d}}{(2 \pi)^{d}} \frac{2 \pi^{\frac{d}{2}}}{\Gamma \left( \frac{d}{2} \right)} \frac{k_{B} T}{\left( \varepsilon_{2,\Lambda} - \mu_{2} \right)^{2}} . \label{eqn:flow3}
\end{align}
Setting $\mu_{1}=\mu_{2}=\mu$ and $V_{0,1}=V_{0,2}=V_{0}$ results in the RG equations for the intra- and inter-component interaction parameters. We rewrite the flow equations in terms of dimensionless variables, which are introduced in the main text:
\begin{align}
&\dfrac{d \mu_{\Lambda}}{dl} = 2 \mu_{\Lambda} - 2 g_{0} \frac{T_{\Lambda}}{1 - \mu_{\Lambda}} - g_{12} \frac{T_{\Lambda}}{1 - \mu_{\Lambda}} , \label{RG1} \\
&\dfrac{d g_{0}}{dl} = \varepsilon g_{0} - 5 g_{0}^{2} \frac{T_{\Lambda}}{\left( 1 - \mu_{\Lambda} \right)^{2}} - g_{12}^{2} \frac{T_{\Lambda}}{\left( 1 - \mu_{\Lambda} \right)^{2}} , \label{RG2} \\
&\dfrac{d g_{12}}{dl} = \varepsilon g_{12} - 2 g_{12}^{2} \frac{T_{\Lambda}}{\left( 1 - \mu_{\Lambda} \right)^{2}} - 4 g_{0} g_{12} \frac{T_{\Lambda}}{\left( 1 - \mu_{\Lambda} \right)^{2}} . \label{RG3}
\end{align}
We determine the  three fixed points of these equations to be $(\mu_{\Lambda}^{\ast}, g_0^{\ast},g^{\ast}_{12})=(0,0,0)$, $(\varepsilon/(5+\varepsilon),5 \varepsilon / ((5+\varepsilon)^{2} T_{\Lambda}),0)$ and $(\varepsilon/(4+\varepsilon),8 \varepsilon / (3 (4+\varepsilon)^{2} T_{\Lambda}),8 \varepsilon / (3 (4+\varepsilon)^{2} T_{\Lambda})$. We Taylor-expand $1/(1 - \mu_{\Lambda})$ such that the corrections in the flow equations are consistently quadratic in the coupling constants \cite{Domany, Rudnick}. Since $\varepsilon_{\Lambda}$ serves as the maximum energy scale in the theory description, i.e. $\mu_{\Lambda}\ll 1$,  the approach is justified, and the approximate RG equations take the form (\ref{eqn:flow-final1-alt}-\ref{eqn:flow-final3-alt}). Compared to the original flow equations (\ref{RG1}-\ref{RG2}), the second and the third fixed point is slightly shifted for these flow equations. However, we find that the nature of the fixed points, including the number of relevant, irrelevant and marginal scaling fields, remains unchanged \cite{Rudnick}. 
 %It implies that the RG flow is qualitatively the same as the flow resulting from the original flow equations.

%%%%%%%%%%%%%%%%%%%
%    Appendix 2    %%%%%%%%%%%%%%%%%%
\section{Expansion around fixed points to first order}\label{flow-first-order}
In this Appendix, we investigate the flow equations (\ref{eqn:flow-final1-alt}-\ref{eqn:flow-final3-alt}) in the vicinity of the fixed points. For that we Taylor-expand the flow equations to first order around them. For each fixed point the three resulting coupled linear differential equations can be solved analytically and the results are sketched qualitatively in Fig. \ref{fig:flow-3d}. The coupled differential equations can be summarized by a $3 \times 3$ matrix. When we diagonalize this matrix, we can identify the so-called scaling fields $v_{\alpha}$ ($\alpha=1,2,3$), directions in the flow diagram that obey a particularly easy relation near the fixed points 
\begin{align}
v_{\alpha} (l) \propto e^{\lambda_{\alpha} l} ,
\end{align}
with the eigenvalues $\lambda_{\alpha}$. When $\lambda_{\alpha} > 0$, the flow will be pushed away from the fixed point along the corresponding direction and the scaling field is said to be relevant. When $\lambda_{\alpha} < 0$, the flow is attracted by the fixed point and the corresponding scaling field is said to be irrelevant. In the case when $\lambda_{\alpha} = 0$, the scaling field is said to be marginal along the corresponding direction and the Taylor expansion up to first order is inconclusive.

For the first (trivial) fixed point (\ref{eqn:fixed-trivial}) Taylor-expansion to first order results in the following differential equations:
\begin{align}
\dfrac{d \Delta \mu_{\Lambda}}{dl} & = 2 \Delta \mu_{\Lambda} - 2 T_{\Lambda} \Delta g_{0} - T_{\Lambda} \Delta g_{12} , \\
\dfrac{d \Delta g_{0}}{dl} & = \varepsilon \Delta g_{0} , \\
\dfrac{d \Delta g_{12}}{dl} & = \varepsilon \Delta g_{12} .
\end{align}
Here, $\Delta \mu_{\Lambda}$, $\Delta g_{0}$ and $\Delta g_{12}$ are the deviations of the coupling constant from their fixed point values $\mu_{\Lambda}^{\ast}$, $g_{0}^{\ast}$ and $g_{12}^{\ast}$:
\begin{align}
\Delta \mu_{\Lambda} & = \mu_{\Lambda} - \mu_{\Lambda}^{\ast} , \\
\Delta g_{0} & = g_{0} - g_{0}^{\ast} , \\
\Delta g_{12} & = g_{12} - g_{12}^{\ast} .
\end{align}
The eigenvalues and the corresponding scaling fields are:
\begin{align}
\lambda_{1}=2, & \quad v_{1} = \Delta \mu_{\Lambda} , \\
\lambda_{2}=\varepsilon, & \quad v_{2} = \frac{2 T_{\Lambda}}{2 - \varepsilon} \Delta \mu_{\Lambda} + \Delta g_{0} , \\
\lambda_{3}=\varepsilon, & \quad v_{3} = \frac{T_{\Lambda}}{2 - \varepsilon} \Delta \mu_{\Lambda} + \Delta g_{12} .
\end{align}
It follows that the first fixed point possesses three relevant scaling fields. Therefore, it is an unstable fixed point pushing the flow away from it.

First-order Taylor-expansion around the second fixed point (\ref{eqn:fixed-old}) results in:
\begin{align} \nonumber
\dfrac{d \Delta \mu_{\Lambda}}{dl} & = 2 \left( 1 - \frac{\varepsilon}{5} \right) \Delta \mu_{\Lambda} - 2 T_{\Lambda} \left( 1 + \frac{\varepsilon}{5} \right) \Delta g_{0} \\
& \hspace{5mm} - T_{\Lambda} \left( 1 + \frac{\varepsilon}{5} \right) \Delta g_{12} , \\
\dfrac{d \Delta g_{0}}{dl} & = - \varepsilon \Delta g_{0} , \\
\dfrac{d \Delta g_{12}}{dl} & = \frac{1}{5} \varepsilon \Delta g_{12} .
\end{align}
\begin{align}
\lambda_{1}=\frac{2(5-\varepsilon)}{5}, & \quad v_{1} = \Delta \mu_{\Lambda} , \\
\lambda_{2}=-\varepsilon, & \quad v_{2} = \frac{2(5+\varepsilon)T_{\Lambda}}{10+3\varepsilon} \Delta \mu_{\Lambda} + \Delta g_{0} , \\
\lambda_{3}=\frac{\varepsilon}{5}, & \quad v_{3} = \frac{(5+\varepsilon)T_{\Lambda}}{10-3\varepsilon} \Delta \mu_{\Lambda} + \Delta g_{12} .
\end{align}
The second fixed point has two relevant and one irrelevant scaling fields. Hence, it is a stable fixed point and the critical surface on which the flow is directed towards it is one-dimensional.

For the third fixed point (\ref{eqn:fixed-new}) first-order expansion gives:
\begin{align} \nonumber
\dfrac{d \Delta \mu_{\Lambda}}{dl} & = 2 \left( 1 - \frac{\varepsilon}{4} \right) \Delta \mu_{\Lambda} - 2 T_{\Lambda} \left( 1 + \frac{\varepsilon}{4} \right) \Delta g_{0} \\
& \hspace{5mm} - T_{\Lambda} \left( 1 + \frac{\varepsilon}{4} \right) \Delta g_{12} , \label{eqn:taylor1-1} \\
\dfrac{d \Delta g_{0}}{dl} & = - \frac{2}{3} \varepsilon \Delta g_{0} - \frac{1}{3} \varepsilon  \Delta g_{12} , \label{eqn:taylor1-2} \\
\dfrac{d \Delta g_{12}}{dl} & = - \frac{2}{3} \varepsilon  \Delta g_{0} - \frac{1}{3} \varepsilon  \Delta g_{12} . \label{eqn:taylor1-3}
\end{align}
\begin{align}
\lambda_{1} = \frac{4-\varepsilon}{2}, & \quad v_{1} = \Delta \mu_{\Lambda} \\
\lambda_{2} = -\varepsilon, & \quad v_{2} = \frac{3 T_{\Lambda}}{2} \Delta \mu_{\Lambda} + \Delta g_{0} + \Delta g_{12} , \\
\lambda_{3} = 0, & \quad v_{3} = - \frac{1}{2} \Delta g_{0} + \Delta g_{12} .
\end{align}
The third fixed point possesses one relevant, one irrelevant and one marginal scaling field. Thus, the third fixed point is stable. The exact nature of the marginal direction of flow and the dimensionality of the critical surface are investigated in the main text.

%\bibliography{references_WM}

\end{document}